\documentclass[utf8]{frontiersFPHY} % for Physics and Applied Mathematics and Statistics articles

\setcitestyle{square} % for Physics and Applied Mathematics and Statistics articles
\usepackage{url,hyperref,lineno,microtype,subcaption}
\usepackage[onehalfspacing]{setspace}
\usepackage{float}
\usepackage{graphicx}
\usepackage{pdfpages}
\def\keyFont{\fontsize{8}{11}\helveticabold }
\def\firstAuthorLast{Kim {et~al.}} %use et al only if is more than 1 author
\def\Authors{Jihwan Kim\,$^{1}$, Do-Hyun Kim\,$^{2,\ast}$, and Jeong-Hyuck Park\,$^{1,\ast}$}
\def\Address{~\\$^{1}$Department of Physics, Sogang University\\  35 Baekbeom-ro, Mapo-gu, Seoul 04107,   Korea \\~\\
$^{2}$Institute for Whole Person Education,  Sogang University\\ 35 Baekbeom-ro, Mapo-gu,  Seoul 04107,   Korea\\~\\~  }
\def\corrEmail{~dohyunkim@sogang.ac.kr   ~\&~  park@sogang.ac.kr}

\begin{document}

%%%%%%%%%%%%%%%%%%%%%%%     Line Spacing   %%%%%%%%%%%%%%%%%%%%%%%
\renewcommand{\baselinestretch}{1.24}   % 1.5 spacing btwn text lines
\setlength{\jot}{8pt}                 % spacing btwn the rows of an eqnarray
\renewcommand{\arraystretch}{2.4}       % spacing btwn the rows of a non-eqn array
%%%%
%
\newcommand{\be}{\begin{equation}}
\newcommand{\ee}{\end{equation} }
\newcommand{\beqa}{\begin{eqnarray}}
\newcommand{\eeqa}{\end{eqnarray} }
\newcommand{\ba}{\begin{array}}
\newcommand{\ea}{\end{array}}
\newcommand\dis{\displaystyle}

\newcommand{\half}{{{\textstyle\frac{1}{2}}}}
\newcommand{\quarter}{{{\textstyle\frac{1}{4}}}}
\newcommand{\su}{\mathbf{su}}
\newcommand{\so}{\mathbf{so}}
\newcommand{\SO}{\mathbf{SO}}
\newcommand{\osp}{\mathbf{osp}}

\newcommand\tr{{\rm tr}}
\newcommand\Tr{{\rm Tr}}

\newcommand\cA{{\cal A}}
\newcommand\cB{{\cal B}}
\newcommand\cC{{\cal C}}
\newcommand\cD{{\cal D}}
\newcommand\cE{{\cal E}}
\newcommand\cG{{\cal G}}
\newcommand\cH{{\cal H}}
\newcommand\cI{{\cal I}}
\newcommand\cJ{{\cal J}}
\newcommand\cL{{\cal L}}
\newcommand\cM{{\cal M}}
\newcommand\cN{{\cal N}}
\newcommand\cO{{\cal O}}
\newcommand\cP{{\cal P}}
\newcommand\cQ{{\cal Q}}
\newcommand\cS{{\cal S}}
\newcommand\cT{{\cal T}}
\newcommand\cV{{\cal V}}
\newcommand\cZ{{\cal Z}}

\newcommand\M{{\Delta}}
\newcommand\rmd{{\rm d}}
\newcommand\rmx{{\rm x}}
\newcommand\rd{{\rm d}}

\newcommand\bfx{{\bf x}}

\newcommand\ti{\tilde{\imath}}
\newcommand\tj{\tilde{\jmath}}

\newcommand\hi{\hat{\imath}}
\newcommand\hj{\hat{\jmath}}
\newcommand\hd{\hat{d}}
\newcommand\hl{\hat{l}}
\newcommand\hn{\hat{n}}
\newcommand\hZ{\hat{Z}}
\newcommand\halpha{{\hat{\alpha}}}
\newcommand\hbeta{{\hat{\beta}}}
\newcommand\talpha{{\tilde{\alpha}}}
\newcommand\tbeta{{\tilde{\beta}}}
\newcommand\tomega{\tilde{\omega}}
\newcommand\homega{\hat{\omega}}
\newcommand\state{{\psi}}

\newcommand\vn{\vec{n}}
\newcommand\db{\td}
\newcommand\td{\tilde{d}}
\newcommand\sD{\scriptstyle{D}}
\newcommand\disOmegaKN{\Omega^{{{{\rm dis.}}}}_{KN}}
\newcommand\idOmegaKN{\Omega^{{{{\rm id.}}}}_{KN}}
\newcommand\dHl{{{\frac{(l+d-1)!}{l!(d-1)!}}}}
\newcommand\rmid{{\rm id.\!}}
\newcommand\rmdis{{\rm dis.\!}}
\newcommand\Par{{\rm P}}
\newcommand\rbox{{\rm box}}
\newcommand\single{\scriptscriptstyle{\!N{=1}}}
\newcommand\kB{k_{{\scriptscriptstyle{\rm B}}}}
\newcommand\leff{l_{{\scriptscriptstyle{\rm eff.}}}}
\newcommand\LdeB{\Lambda_{{\scriptscriptstyle{\rm de\,Broglie}}}}
\newcommand\conti{{{\scriptscriptstyle{\rm BEC}}}}
\newcommand\superc{{{\scriptscriptstyle{\rm supercool\,}}}}
\newcommand\superh{{{\scriptscriptstyle{\rm superheat}}}}
\newcommand\Zcf{\mathfrak{Z}_{\scriptscriptstyle{\rm C.F.}}}
\newcommand\fst{1}
%\newcommand\fst{1{\rm st}}

% COLORS ---------------------------------------------------------
%\definecolor{green}{rgb}{0.5,1,0.5}
\newcommand{\red}[1]{{\color{red} #1 \color{black}}}

\newcommand{\blue}[1]{{\color{blue} #1 \color{black}}}

\newcommand{\darkblue}[1]{\textcolor[rgb]{0.1,0.1,0.7}{#1}}

\newcommand{\purple}[1]{\textcolor[rgb]{0.8,0,1}{#1}}

\newcommand{\orange}[1]{{\color{orange} #1 \color{black}}}
\newcommand{\sepia}[1]{{\color{sepia} #1 \color{black}}}

\newcommand{\green}[1]{\textcolor[rgb]{0.2,0.5,0.1}{#1}}

\onecolumn
\firstpage{1}
\title[Janus van der Waals equations   for real molecules  with  two-sided phase transitions]{~\\~\\Janus van der Waals equations  for real molecules  with  two-sided phase transitions\\~\\~} 

\author[\firstAuthorLast ]{\Authors} %This field will be automatically populated
\address{} %This field will be automatically populated

\correspondance{} %This field will be automatically populated

\extraAuth{}% If there are more than 1 corresponding author, comment this line and uncomment the next one.

\maketitle

\begin{abstract}
\noindent We obtain  families  of generalised  van der Waals equations   characterised by an even number ${n=2,4,6}$ and  a  continuous  free parameter which is tunable   for a  critical compressibility factor. Each equation   features two adjacent  critical points  which have  a  common critical  temperature $T_{c}$  and arbitrarily close two critical  densities. The critical phase transitions are naturally two-sided: the critical  exponents  are  $\alpha_{\scriptscriptstyle{P}}=\gamma_{\scriptscriptstyle{P}}=\frac{2}{3}$, $\beta_{\scriptscriptstyle{P}}=\delta^{-1}=\frac{1}{3}$  for $T>T_{c}$  and  $\alpha_{\scriptscriptstyle{P}}=\gamma_{\scriptscriptstyle{P}}=\frac{n}{n+1}$, $\beta_{\scriptscriptstyle{P}}=\delta^{-1}=\frac{1}{n+1}$ for $T<T_{c}$.  In  contrast with the original van der Waals equation,   our novel equations  all reduce  consistently   to the classical ideal gas law in low density limit.   We test our formulas against NIST  data  for eleven major molecules   and show    agreements better than the original van der Waals equation, not only near to  the critical points but also in low density regions.

\tiny
 \keyFont{ \section{Keywords:} Equations of State, van der Waals Equation, Critical Point, Two-sided Phase Transition,   NIST Reference Data, Analyticity} \hfill%All article types: you may provide up to 8 keywords; at least 5 are mandatory.
\end{abstract}
\vspace{5pt}
%\newpage
\section{Introduction}
Two-sided phase transitions are  rather out of the ordinary    critical phenomena as their  critical exponents take  different values in the higher and the lower temperature phases. While  the general renormalization group argument, \textit{e.g.}~\cite{Justin}, might  appear to suppress     such an unusual bilateral   critical behaviour, they have  been       reported to  occur   in various systems, such as  isotropic ferromagnet~\cite{Nelson},  XY-Heisenberg model~\cite{Leonard:2015wyg}, complex  Sachdev-Ye-Kitaev models~\cite{Azeyanagi:2017drg}, and  liquid-gas transitions of real molecules~\cite{Cho:2016jzz}.  To explain  the two-sided critical phase transitions, while respecting  the  analyticity of the canonical  partition function of a finite system, it was hypothesized    that there may exist not a single but double  critical points which should be  quite  close to each other~\cite{Cho:2016jzz}.

It is the dual-purpose of the present paper   to  modifiy  the  van der Waals equation  toward the description of the two-sided phase transitions  and   to test  our novel formulas  against NIST  Reference Data (RRID:SCR$\underline{~~}$006452)~\cite{NIST},  specifically   for {eleven} real molecules.   Our proposed  equations of state, which we dub  \textit{Janus van der Waals equations},\footnote{`Janus' is a Roman god who has two faces. Our nomenclature is  inspired partially  by \cite{Bak:2003jk}.}   are characterised by   extremely   adjacent   two critical points. The two critical points  share  strictly  the same critical temperature, $T_{c}$, but,  remarkably,   the critical pressure, $P_{c}$, and the critical  volume per particles, $v_{c}$,    differ by  arbitrarily small amounts, such that  the disticnt  two critical points can appear practially indistinguishable.

The  (original) van der Waals equation of state,
\be
\left(P_{r}+\frac{3}{v_{r}^{2}}\right)\left(v_{r}-\frac{1}{3}\right)=\frac{8}{3}T_{r}\,,
\label{vdW}
\ee
was  meant to be an improvement of the classical ideal gas law, \textit{i.e.~}
\be
	Pv=\kB T\qquad\Longleftrightarrow\qquad P_{r}v_{r}=\chi T_{r}\quad:\quad
	\chi=\frac{\kB T_{c}}{P_{c}v_{c}}\,,
\label{ideal}
\ee
by attempting  to take into account   the  finite volume of molecules and intermolecular attractions. In (\ref{vdW}), (\ref{ideal}), and henceforth,  $P_{r}=P/P_{c}$,  $T_{r}=T/T_{c}$, and $v_{r}=v/ v_{c}$  are the  reduced pressure, temperature, and  volume per particle respectively, while $\chi={\kB T_{c}}/({P_{c}v_{c}})$  denotes the  inverse of the critical compressibility factor which is dimensionless.  The critical point is   given by  $P_{r}=T_{r}=v_{r}=1$, such that, the van der Waals equation~(\ref{vdW}) contains the critical point as $(1+3)(1-\frac{1}{3})=\frac{8}{3}$, and may describe the near  critical behaviour.  However,  in a  large volume or low density limit,  Eq.(\ref{vdW})	gives
\be
Pv\simeq\left(\frac{8P_{c}v_{c}}{3T_{c}}\right)T\neq\kB T\,.
\label{asymptotic}
\ee
Thus, unless  $\chi=\frac{8}{3}$ by chance,  the van der Waals equation~(\ref{vdW})  cannot reduce to the classical ideal gas law~(\ref{ideal}) in the low density limit, and accordingly  fails to describe real gases at  low densities.  In fact,   experimental real values of $\chi$ are   typically around $3.5$  larger  than $\frac{8}{3}$.  In contrast, our proposed Janus  van der Waals equations are    going to be consistent with  the classical ideal gas law~(\ref{ideal})  at  low densities.  They  are not  particularly motivated by the finite volume or intermolecular effects. Rather, we    address  directly  the  definition of the critical point in thermodynamics: for a given equation of state,   a critical point can be identified as a stationary inflection point in the constant temperature line on a pressure \textit{versus} volume diagram. Specifically for a certain natural number  greater than or equal to two, $n_{c}\geq 2$, we have at the critical point,
\be
\ba{lll}
\dis{\frac{\partial^{k}P(T_{c},v_{c})}{\partial v^{k}}=0}~~&~\mbox{for}~&~~1\leq k\leq n_{c}\,,
\ea
\label{defc}
\ee
while the next higher order derivative having  ${k={n_{c}+1}}$ is nontrivial.  In particular for the  van der Waals equation~(\ref{vdW}),  the  number  takes the minimal value ${n_{c}=2}$ and  its spinodal curve that is by definition the lowest order  ${k=1}$ in (\ref{defc})  is given by
\be
T_{r}=1-\frac{(v_{r}-1)^{2}(4v_{r}-1)}{4v_{r}^{3}}\,.
\label{spinodalvdW}
\ee
This expression shows clearly  that   on the spinodal curve the temperature is locally maximal at the   critical point.  In general,  the characteristic  number $n_{c}$ can differ from two and  may be used to   classify the critical points, being  dubbed  as `\textit{critical  index}'~\cite{Cho:2016jzz}.  Thoroughly from  (\ref{defc}),   both  the critical isobar of  $P_{r}\equiv 1$ and   the spinodal curve  satisfy simple power-law behaviors  around the critical point,% on the temperature-volume plane, 
\be
\ba{ll}
T-T_{c}~\propto~(v-v_{c})^{1+n_{c}}~:~\mbox{critical~isobar}\,,\quad~&~\quad
T-T_{c}~\propto~(v-v_{c})^{n_{c}}~:~\mbox{spinodal~curve}\,.
\ea
\label{np1}
\ee 
Combined with the analyticity of the underlying canonical partition function,  this result  fixes   the (isobaric)  critical exponents~\cite{7616relativistic},
$\alpha_{\scriptscriptstyle{P}}=\gamma_{\scriptscriptstyle{P}}=\frac{n_{c}}{\,1+n_{c}}$ and $\beta_{\scriptscriptstyle{P}}=\delta^{-1}=\frac{1}{\,1+n_{c}}$. These  satisfy     Rushbrooke and  Widom scaling laws. Since $n_{c}\geq 2$, the two curves of (\ref{np1})  are actually  tangent to each other at the critical point.

 The  main result of \cite{Cho:2016jzz} was  that  the NIST Reference Data of  twenty major molecules~\cite{NIST} are  indeed  consistent with the analytic prediction of the critical exponents, and moreover that  the critical phase transitions are remarkably    two-sided:   for $T>T_{c}$  the critical index is $ 2$   universally, yet   for $T<T_{c}$   it varies as  $n\equiv 2,3,4,5,6$  depending on each molecule, collectively denoted by a pair of critical indices, $n_{c}\,\Rightarrow\,(n_{+},n_{-})=(2,n)$.  In this work, we present  Janus van der Waals equations  characterised by  a pair of  adjacent critical points with  indices $(n_{+},n_{-})=(2,n)$ for  $n=2,4,6$ (even). With one continuous input parameter  which will be   chosen  to match the critical compressibility,   they   are shown to   describe  remarkably  well   all the  real  molecules with even critical indices identified  in \cite{Cho:2016jzz}.  They are  cyclopentane~($\mathrm{C}_{5}\mathrm{H}_{10}$) for ${n=2}$;  nitrogen~($\mathrm{N}_{2}$), argon~($\mathrm{Ar}$), methane~($\mathrm{CH}_{4}$), ethylene~($\mathrm{C}_{2}\mathrm{H}_{4}$), ethane~($\mathrm{C}_{2}\mathrm{H}_{6}$),  propylene~($\mathrm{C}_{3}\mathrm{H}_{6}$), propane~($\mathrm{C}_{3}\mathrm{H}_{8}$), butane~($\mathrm{C}_{4}\mathrm{H}_{10}$),  and isobutane~($\mathrm{C}_{4}\mathrm{H}_{10}$) for ${n=4}$; and helium-4~(${}^{4}\mathrm{He}$) for ${n=6}$.  

\section{Ansatz and Derivation}\vspace{-12pt}
Our Janus van der Waals equations   assume an ansatz,
\be
\Big(P_{r}+\chi f_{n}(v_{r})\Big)\left(v_{r}-b\right)=\chi\, T_{r}\,,
\label{bcvdW}
\ee
where    $f_{n}(v_{r})$ is  supposed to be a polynomial in $v_{r}^{-1}$ and $\chi=\frac{\kB T_{c}}{P_{c}v_{c}}$ is  the (experimentally determinable) genuine   \textit{free} parameter which will guarantee  the consistency with the classical ideal gas law~(\ref{ideal}) in the  large volume limit, hence resolving  the inconsistency of the original van der Waals equation~(\ref{asymptotic}).  Each molecule will  have  its own  Janus  van der Waals equation characterised by two input parameters,  $\chi$ (continuous) and $n=2,4,6$ (discrete).

\begin{figure}[H]
\begin{center}
\vspace{-5pt}
%\begin{minipage}{.5\textwidth}
%\end{minipage}
%\,\qquad\,
%\begin{minipage}{.4\textwidth}
\includegraphics[width=8.50cm]{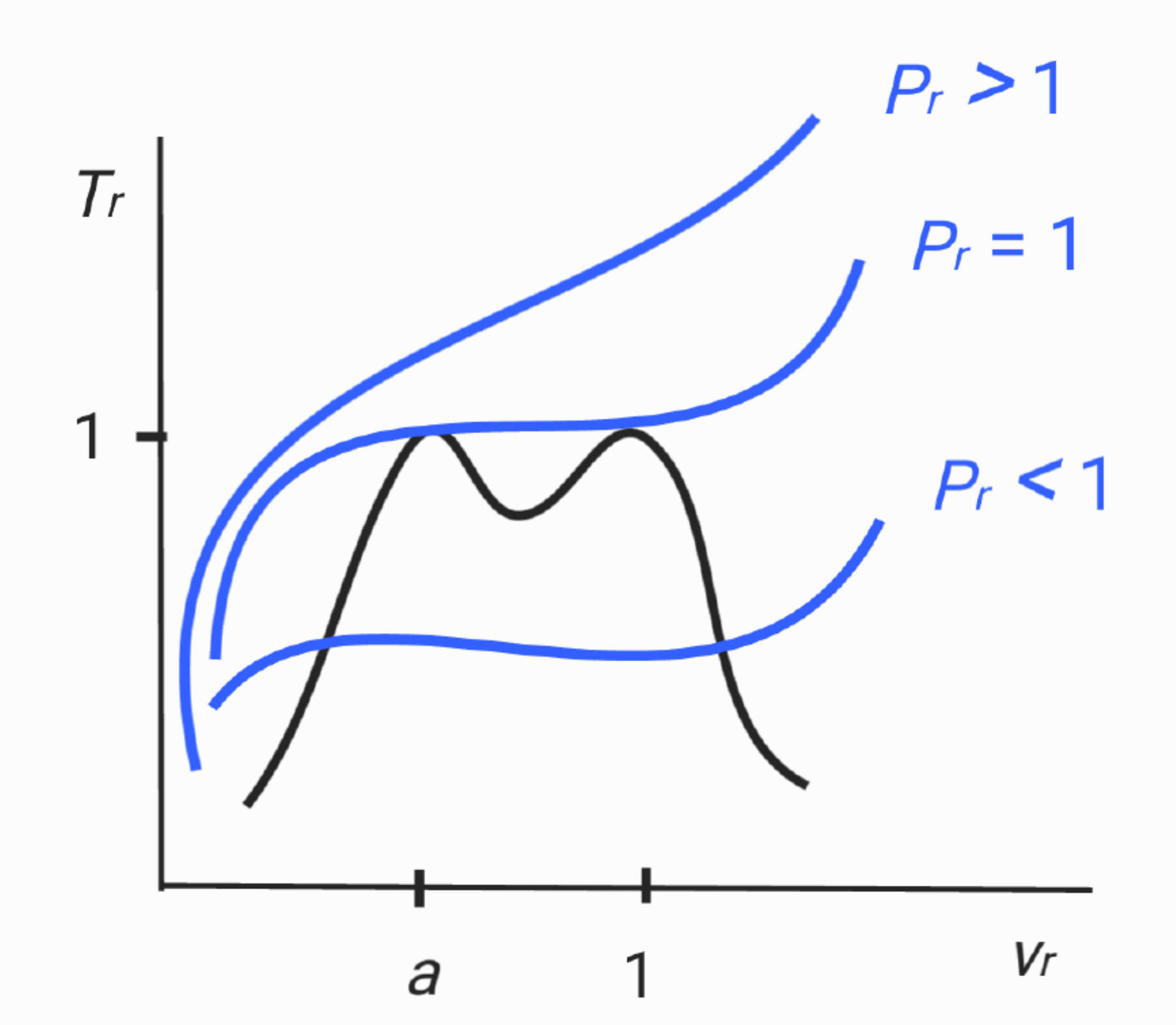}
\caption{Schematic zoomed-in diagram of a spinodal curve (black colored) and three isobar curves \blue{(blue colored)}.  There are two adjacent critical points on a spinodal curve at $v_{r}=a$ and at $v_{r}=1$.   When $P_{r}<1$ the isobar crosses the spinodal curve twice at $v_{r}<a$ and at $v_{r}>1$.  When $P_{r}>1$ the isobar does not meet the spinodal curve.  When $P_{r}=1$, with negligible $\epsilon$, the critical isobar touches the spinodal curve at the two critical points. Specifically, along the critical isobar,   as  the temperature increases from $T_{r}<1$  to $T_{r}=1^{-}$, the critical point at $v_{r}=a$ with the critical index of $n_{c}=n=2,4,6$ is relevant. On the other hand,   as the temperature decreases from $T_{r}>1$  to $T_{r}=1^{+}$, the dominant  critical point is at $v_{r}=1$ with $n_{c}=2$.\\
\mbox{~~~~}Between the two peaks on the spinodal curve, there is a local minimum  which may be also identified as a    ``critical point"   of which the critical temperature is less than $T_{c}$.   This third critical point is sandwiched by  the two extremely adjacent critical points and  plays little  role in our analysis.   The  diagram generalises the single-peaked spinodal curve of   a relativistic ideal Bose gas, depicted in \textbf{Figure 1 of Ref.\cite{7616relativistic}}, to twin-peaks, and furthermore    will be confirmed   through a concrete example  as in \textbf{Figure~\ref{FIG2}}.}\label{FIG1}
%\end{minipage}
\end{center}
\end{figure}

With four constants $\{a,b,s,t\}$ which will be determined shortly,      we require the spinodal curve to meet   
\be
T_{r}=\dis{-(v_{r}-b)^{2}\frac{\rd f_{n}(v_{r})}{\rd v_{r}}}
=\dis{1-\frac{(v_{r}-a)^{n}(v_{r}-1)^{2}(v_{r}^{2}-sv_{r}+t)}{v_{r}^{n+4}}\,.}
\label{spinodalT}
\ee
While the first equality comes from the definition of the spinodal curve,  crucially  the second  generalises  the  van der Waals case~(\ref{spinodalvdW}) and  gives rise to   two distinct critical points:
\be
\ba{lll}     
\!(n_{c},P_{r},T_{r},v_{r})\,=\, (n,1+\epsilon,1,a) &~\mbox{and}~& (2,1,1,1)\,.
\ea
\label{twoc}
\ee
At each critical point, in view of (\ref{np1}), we have clearly
\be
T_{r}-1\propto\left\{\ba{lll}(v_{r}-a)^{n}&\mbox{as}&v_{r}\,\rightarrow\, a\\
(v_{r}-1)^{2}&\mbox{as}&v_{r}\,\rightarrow \,1\ea\right.\,.
\ee
Their critical indices, pressure, and volume per particle may  differ, but the critical temperature is the same, \textit{i.e.~}${T_{r}=1}$.  
For this,  it is necessary  to set  ${0<a<1}$ and ${s^{2}\neq 4t}$. We also put $n$ to be even, $n=0,2,4,6,\cdots$,  such that the critical temperature is (locally) maximal  on the spinodal curve, which is the case with  the  van der Waals fluid~(\ref{spinodalvdW}), a  relativistic ideal Bose gas~\cite{7616relativistic} (Figures 2 and 6 therein),  and supposedly real molecules. Accordingly  the spinodal curve~(\ref{spinodalT}) has `twin peaks' in temperature,  as depicted  in  \textbf{Figure~\ref{FIG1}}.  Moreover, in order to be consistent with the realistic molecules for which only one critical  point has been usually assumed,  we shall let $a$ be close to $1^{-}$ and then,  from $\epsilon\propto(1-a)^{n+3}$ which we shall derive later in (\ref{epsilon}), the two critical pressure, $P_{r}=1$ and $P_{r}=1+\epsilon$, will be practically indistinguishable from each other.

From the fact that the last expression  in (\ref{spinodalT}) should be divisible by $(v_{r}-b)^{2}$,  two constraints arise:
\be
\ba{ll}
(b-1)^{2}(b-a)^{n}(b^{2}-sb+t)=b^{n+4}\,,\qquad&\qquad
\frac{2}{b-1}+\frac{n}{b-a}+\frac{2b-s}{b^{2}-sb+t}=\frac{n+4}{b}\,,
\ea
\label{divisible} 
\ee
which  subsequently  determine    $s$ and $t$ in terms of $a$ and $b$,
\be
\ba{ll}
s=2b+\left(\frac{\,(n+2)ab-(n+4)a+4b-2b^{2}\,}{(b-1)^{3}(b-a)^{n+1}}\right)b^{n+3}\,,\qquad&\qquad
t=b^{2}+\left(\frac{\,(n+1)ab-(n+3)a+3b-b^{2}\,}{(b-1)^{3}(b-a)^{n+1}}\right)b^{n+4}\,.
\ea
\label{u12}
\ee
With these identifications, the second equality of (\ref{spinodalT}) fixes the function $f_{n}(v_{r})$,  
\be
f_{n}(v_{r})\,=\,\frac{1}{b^{5}}\,\sum_{l=0}^{n+1}\,\frac{c_{l}}{n+3-l}\left(\frac{b}{v_{r}}\right)^{n+3-l}\,,
\label{fform}
\ee
where the coefficients are given, with (\ref{u12}), by\footnote{Essentially, the coefficients~(\ref{arec}) stem from  a     recurrence relation  $c_{l}-2c_{l-1}+c_{l-2}=h_{l}$ for some $h_{l}$, whose   solution reads in general  $c_{l}=\sum_{j=0}^{l} (l+1-j)h_{j}$.} %$\{a,b\}$  through
\be
c_{l}={\dis{\sum_{j=0}^{l}}}~{{(j{-l}{-1})}}
\Big[\textstyle{\binom{n}{j-4\,}a^{4}
+\binom{n}{j-3\,}(2{+s})a^{3}
+\binom{n}{j-2\,}(1{+2s}{+t})a^{2}
+\binom{n}{j-1\,}(s{+2t})a
+\binom{n}{j} t}
\Big]\!\left(-\frac{a}{b}\right)^{n-j}\,.
\label{arec}
\ee
Note that   the  binomial  coefficient $\binom{n}{k}=\frac{n!}{k!(n-k)!}$  should  be trivial if  $k$ or $n-k$ is negative.

The remaining two constants $a,b$ are  then determined  by requiring  that the reduced critical pressure  $P_{r}$ should take the  aforementioned  values of  $1+\epsilon$ and $1$  at the two critical points of ${v_{r}=a}$ and ${v_{r}=1}$~(\ref{twoc}):  
\be
\ba{ll}
\chi\left[\frac{1}{a-b}-f_{n}(a)\right]=1+\epsilon\,,\qquad&\qquad \chi\left[\frac{1}{1-b}-f_{n}(1)\right]=1\,.
\ea
\label{P1}
\ee
We obtain from the latter 
\be
\chi=\frac{(n+3)(1-b)^{n+3}}{(1-b)^{n+3}+b^{n+3}}\,,
\label{chib}
\ee
and subtracting the latter from the former
\be
\epsilon= 
\frac{2\chi(1-a)^{n+3}\left[(1-b)^{3}(a-b)^{n+1}+
\left\{(n+1)a-b+3\right\}b^{n+3}\right]}{(n+3)(n+2)(n+1)a^{3}(1-b)^{3}(a-b)^{n+1}}\,.
\label{epsilon}
\ee
Inverting (\ref{chib}), we  solve for $b$  in terms of  the physically measurable compressibility factor,  
\be
b={\frac{(n+3-\chi)^{\frac{1}{n+3}}}{
(n+3-\chi)^{\frac{1}{n+3}}+\chi^{\frac{1}{n+3}}}}\,.
\label{b}
\ee
%%
%Subsequently,  substituting  this into (\ref{epsilon}),   we can express $\epsilon$ in terms $\chi$ as well,
%\be
%\epsilon\red{=???}
%\frac{2(1-a)^{n+3}}{(n+2)(n+1)}\left[n+4-\chi+(n+3-\chi)\left(\frac{n+3-\chi}{\chi}\right)^{\frac{1}{n+3}}\right]\,.
%\ee
%%
Clearly from (\ref{epsilon}),  $\epsilon$ becomes  small   $\epsilon\propto({1-a})^{n+3}$ as  the constant $a$ gets close to unity  from below. In fact, $\epsilon$ is positive when ${0<a<1}$ and ${0<\chi<n+3}$.  This confirms that  the two critical points~(\ref{twoc}) can be indeed  extremely adjacent and  experimentally indistinguishable.  Naturally,  the limit  $a\rightarrow 1^{-}$   does not  match  the exact  value of ${a=1}$:  the former is still bi-critical while  the latter is mono-critical with the enhanced  critical index ${n_{c}=n{+2}}$.  However,   away from the critical points in the phase diagram, we may   practically   put  
\be
{a\approx 1}\,,
\label{aapprox}
\ee 
and obtain  approximate Janus van der Waals equations which we shall test against NIST Reference Data~\cite{NIST}. 

\begin{figure}[H]
	\begin{center}%\vspace{-24pt}
		\includegraphics[width=19cm]{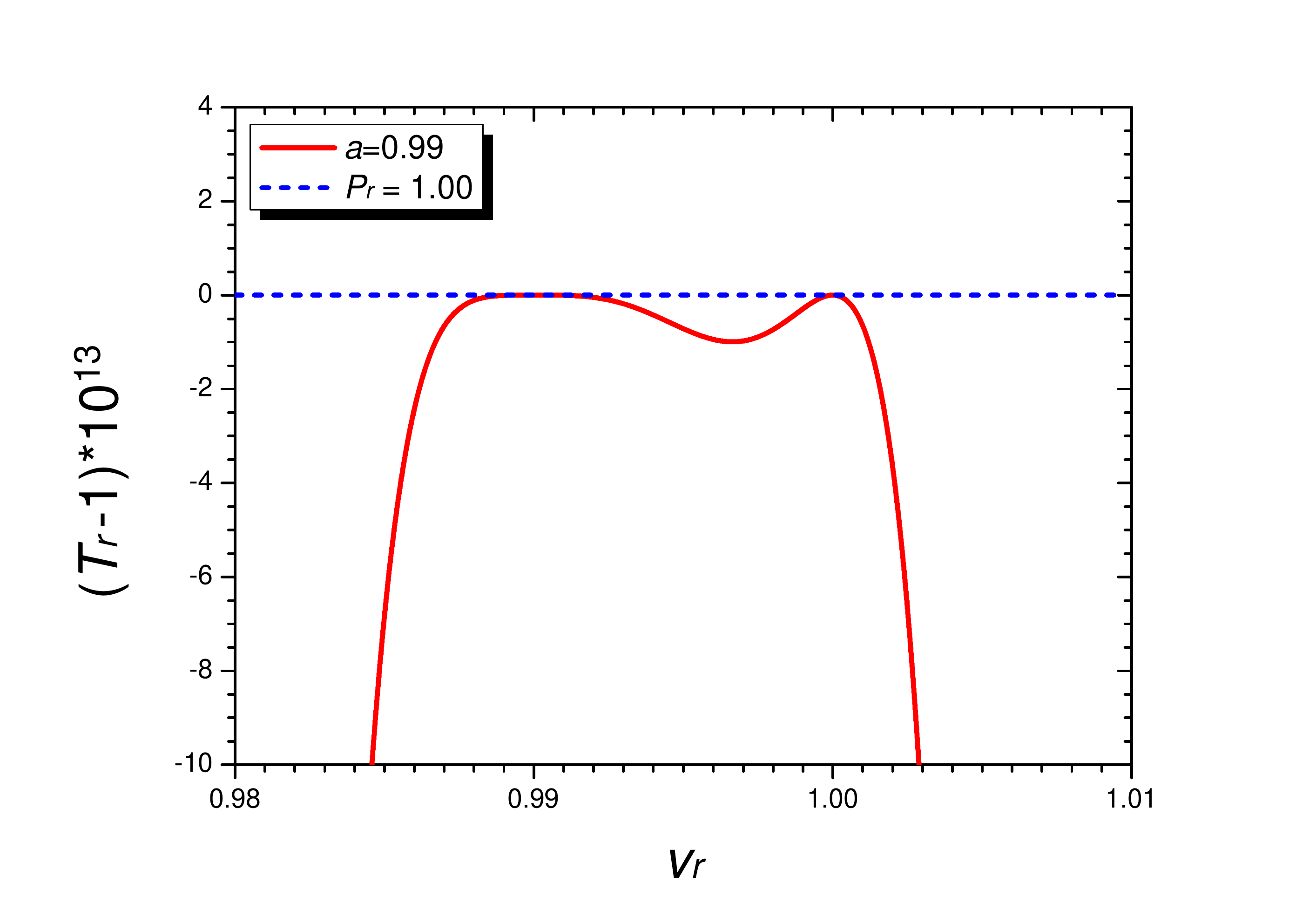}
		\caption{Diagram featuring the spinodal curve \red{(red colored)}  and the critical isobar \blue{(blue colored)} of  the $n=4$ Janus van der Waals equation with   two adjacent critical points at $v_{r}=a=0.99$ and $v_{r}=1.00$. We have set $\chi=3.4556$ as for nitrogen $(\mathrm{N}_{2})$.  Since $\epsilon = 6.0846\times10^{-15}$~(\ref{epsilon}), the two critical isobars, $P_{r}=1$ and $P_{r}=1+\epsilon$, are indeed practically indistinguishable.   The (common)  critical isobar then touches  both   critical points, as anticipated in  \textbf{Figure~\ref{FIG1}}.  The local minimum value of $T_{r}$ sandwiched by the twin critical points is  lower than $1$ by  tiny  small amount $(< 10^{-12})$, and thus also  experimentally hard  to detect. }\label{FIG2}\vspace{-24pt}
	\end{center}
\end{figure}

\textbf{Figure~\ref{FIG2}} is the diagram of a spinodal curve (\ref{spinodalT}) for the choice of ${n=4}, {a=0.99}$, and $\chi = 3.4556$ as for nitrogen ($\mathrm{N}_{2}$). It follows from (\ref{b}) $b=0.5009$ and subsequently   (\ref{u12}) fixes $s, t$. The curve confirms the anticipated   two adjacent critical points at $a=0.99$ and $1.00$.   Spinodal curves for  other molecules can be obtained by the same method.  

\newpage

%%%%%%%%%%%%%%%%%%%%%%%%%%%%%%%%%%%%
%%%%%%%%%%%%%%%%%%%%%%%%%%%%%%%%%%%%

\section{Theoretical result: Janus van der Waals equations}
We  now spell our modified   van der Waals equations  for each case of $n=0$ (mono-critical) and $n=2,4$ (bi-critical/Janus) explicitly.

%%%%%%%%%%%%%%%%%%%%%%%%%%%%%%%%%%%%

\subsection{$n=0$ : mono-critical generalisation}\vspace{-12pt}
When $n=0$, the parameter $a$ becomes irrelevant as  there is  only one critical point. Consequently,  (\ref{bcvdW}) reduces to an one-parameter generalisation of the van der Waals equation,
\be
\left[P_{r}+
\frac{{{\left[\chi^{\frac{1}{3}\!}-(\chi-3)^{\frac{1}{3}\!}\right]\!\left[\chi^{\frac{2}{3}\!}-\left|\chi-3\right|^{\frac{2}{3}\!}\right]}}}{v_{r}^{2}}
+\frac{{{2-\chi+\chi^{\frac{1}{3}\!}\left|\chi-3\right|^{\frac{2}{3}\!}}}}{v_{r}^{3}}
\right]\!\left[v_{r}-\frac{(\chi-3)^{\frac{1}{3}}}{(\chi-3)^{\frac{1}{3}}-\chi^{\frac{1}{3}}}\right]=\chi\, T_{r}\,.
\label{bcvdWn03}
\ee
The critical compressibility factor  is at our disposal.  Choosing $\chi=\frac{\kB T_{c}}{P_{c}v_{c}}=\frac{8}{3}$, one  recovers   the original van der Waals equation~(\ref{vdW}). 

Among the eleven major molecules we examine, cyclopentane is  exceptional, as its critical phase transition is not two-sided: ${n_{c}=2}$  universally for the temperature ${T>T_{c}}$ and ${T<T_{c}}$~\cite{Cho:2016jzz}. Logically, it can be either mono-critical with    ${n=0}$ or  bi-critical with ${n=2}$. To examine  which is correct  with     the NIST data, we put $\chi=3.5572$ as for the value of cyclopentane,  and prepare an ${n=0}$ equation from (\ref{bcvdWn03}), 
\be
\left(P_{r}+
\frac{1.1632}{
v_{r}^{2}}-\frac{0.52356}{v_{r}^{3}}
\right)\!\left(v_{r}+1.1694\right)= 3.5572\, T_{r}\,.
\label{n0JvdW}
\ee 
%%%
%%\be
%%\small{
%%\left(P_{r}+
%%\frac{\,1.15\,}{\,
%%v_{r}^{2}\,}-\frac{\,0.519\,}{\,v_{r}^{3}\,}
%%\right)\!\left(v_{r}+1.19\right)\approx 3.57\, T_{r}\,.}
%%\label{bcvdWC5H10}
%%\ee 
%%%
The plus sign in the second bracket (which is generically the case for $\chi>3)$ contrasts with the negative sign in the original van der Waals equation~(\ref{vdW}).

%%%%%%%%%%%%%%%%%%%%%%%%%%%%%%%%%%%%

\subsection{${n=2}$ equation for cyclopentane~($\mathrm{C}_{5}\mathrm{H}_{10}$)}\vspace{-12pt}
For  ${n=2}$, in the limit ${a\rightarrow 1^{-}}$ or ${a\approx 1}$~(\ref{aapprox}),  we have
\be
\ba{lll}
\chi f_{n=2}(v_{r})\!\!\!&\approx\!\!\!&
\frac{\chi^{\frac{1}{5}}(5-\chi)^{\frac{1}{5}}
\left[\chi^{\frac{1}{5}}-(5-\chi)^{\frac{1}{5}}\right]\!\left[\chi^{\frac{2}{5}}+(5-\chi)^{\frac{2}{5}\!}\right]+4\chi-10\,}{v_{r}^{2}}\,+\,\frac{\chi^{\frac{1}{5}}(5-\chi)^{\frac{2}{5}}
\left[\chi^{\frac{2}{5}}-2\chi^{\frac{1}{5}}(5-\chi)^{\frac{1}{5}}+3(5-\chi)^{\frac{2}{5}}\right]
-6\chi+20\,}{v_{r}^{3}}\\
{}&{}&+\,\frac{\chi^{\frac{1}{5}}(5-\chi)^{\frac{3}{5}}
\left[\chi^{\frac{1}{5}}-3(5-\chi)^{\frac{1}{5}}\right]
+4\chi-15\,}{v_{r}^{4}}\,+\,\frac{\chi^{\frac{1}{5}}(5-\chi)^{\frac{4}{5}}
-\chi+4\,}{v_{r}^{5}}\,.
\ea
\ee
Letting  ${\chi=3.5572}$,  we obtain an ${n=2}$ Janus van der Waals equation for $\mathrm{C}_{5}\mathrm{H}_{10}$ (cyclopentane), 
\be
\left(P_{r}+
\frac{5.0608}{
v_{r}^{2}}+\frac{2.1811}{v_{r}^{3}}-
\frac{3.8860}{v_{r}^{4}}+\frac{2.1710}{v_{r}^{5}}
\right)\!\left(v_{r}-0.45500\right)\approx 3.5572\, T_{r}\,.
\label{n2JvdW}
\ee 
As we shall see in the next section, the NIST data of cyclopentane is better described by this bi-critical equation rather than the mono-critical one~(\ref{n0JvdW}).

%%%%%%%%%%%%%%%%%%%%%%%%%%%%%%%%%%%%

\subsection{${n=4}$ equations  for nitrogen~($\mathrm{N}_{2}$), argon~($\mathrm{Ar}$), methane~($\mathrm{CH}_{4}$), ethylene~($\mathrm{C}_{2}\mathrm{H}_{4}$), ethane~($\mathrm{C}_{2}\mathrm{H}_{6}$),  propylene~($\mathrm{C}_{3}\mathrm{H}_{6}$), propane~($\mathrm{C}_{3}\mathrm{H}_{8}$), butane~($\mathrm{C}_{4}\mathrm{H}_{10}$), and  isobutane~($\mathrm{C}_{4}\mathrm{H}_{10}$)}\vspace{-12pt}
For ${n=4}$ with $b={\frac{(7-\chi)^{\frac{1}{7}}}{
(7-\chi)^{\frac{1}{7}}+\chi^{\frac{1}{7}}}}$ and  ${a\approx 1}$~(\ref{aapprox}),  we have
\be
\ba{lll}
\chi f_{n=4}(v_{r})\!\!\!&\approx\!\!\!&
\frac{21-140 b+392 b^2-588 b^3+490 b^4-196 b^5}{\left[b^7+(1-b)^7\right]v_{r}^{2}}\,+\,\frac{-35+245 b-728 b^2+1176 b^3-1078 b^4+490 b^5}{\left[b^7+(1-b)^7\right]v_{r}^{3}}\\
{}&{}&+\,\frac{35-245 b+735 b^2-1218 b^3+1176 b^4-588 b^5}{\left[b^7+(1-b)^7\right]v_{r}^4}\,+\,\frac{-21+147 b-441 b^2+735 b^3-728 b^4+392 b^5}{\left[b^7+(1-b)^7\right]v_{r}^5}\\
{}&{}&+\,\frac{7-49 b+147 b^2-245 b^3+245 b^4-140 b^5}{\left[b^7+(1-b)^7\right]v_{r}^6}\,+\,\frac{-1+7 b-21 b^2+35 b^3-35 b^4+21 b^5}{\left[b^7+(1-b)^7\right]v_{r}^7}\,.
\ea
\ee
Choosing  $\chi$ from experimental data,  we obtain ${n=4}$ Janus van der Waals equations  for nine molecules,
% nitrogen, argon, methane, ethylene, ethane,  propylene, propane, butane, and isobutane respectively,
 \be
\left(P_{r}+
\frac{k_{2}}{v_{r}^{2}}
+\frac{k_{3}}{v_{r}^{3}}
+\frac{k_{4}}{v_{r}^{4}}+
\frac{k_{5}}{v_{r}^{5}}+
\frac{k_{6}}{v_{r}^{6}}+
\frac{k_{7}}{v_{r}^{7}}
\right)\!\left(v_{r}-b\right)\approx \chi T_{r}\,,
\label{n4JvdW}
\ee 
of which the coefficients are listed in {\textbf{Table~\ref{TableCoeff}}}. \vspace{-3pt}
%\begin{center}
\begin{table}[h]
\scalebox{1}{
\begin{tabular}{|r|cccccccc|}
\hline
~~molecule (${n=4}$) &$\chi$&$k_{2}$&$k_{3}$&$k_{4}$&$k_{5}$&$k_{6}$&$k_{7}$&$b$\\
\hline
~~~nitrogen ~($\mathrm{N}_{2}$) &~3.4556&-0.30474&28.762&-57.117&56.913&-28.406&6.0760&0.50091~\,\\
~~~~argon ~($\mathrm{Ar}$) &3.4542&-0.31380&28.784&-57.150&56.941&-28.418&6.0783&0.50093~\,\\
~~methane ~($\mathrm{CH}_{4}$) &3.4936&-0.044104&28.110&-56.162&56.132&-28.059&6.0110&0.50013~\\
~ethylene ~($\mathrm{C}_{2}\mathrm{H}_{4}$) &3.5563&0.38638&27.034&-54.582&54.840&-27.484&5.9032&0.49885~\,\\
~~ethane ~($\mathrm{C}_{2}\mathrm{H}_{6}$) &3.5726&0.49759&26.755&-54.174&54.505&-27.335&5.8752&0.49852~\,\\
~propylene ~($\mathrm{C}_{3}\mathrm{H}_{6}$) &3.6279&0.87670&25.806&-52.781&53.364&-26.827&5.7797&0.49739~\,\\
~~propane ~($\mathrm{C}_{3}\mathrm{H}_{8}$) &3.6168&0.80075&25.996&-53.060&53.593&-26.929&5.7989&0.49762~\,\\
~~~butane ~($\mathrm{C}_{4}\mathrm{H}_{10}$) &3.6529&1.0482&25.376&-52.150&52.847&-26.597&5.7363&0.49688~\,\\
~isobutane ~($\mathrm{C}_{4}\mathrm{H}_{10}$) &3.6251&0.85816&25.852&-52.849&53.420&-26.852&5.7844&0.49744~\,\\
\hline
\end{tabular}}
\caption{The coefficients of the ${n=4}$ Janus van der Waals equations for nine molecules.}
    \label{TableCoeff}
\end{table}
%\end{center}

\subsection{${n=6}$ equation  for helium-4 (${}^{4}\mathrm{He}$) }

For  ${n=6}$ with  $b={\frac{(9-\chi)^{\frac{1}{9}}}{
		(9-\chi)^{\frac{1}{9}}+\chi^{\frac{1}{9}}}}$  and  ${a\approx 1}$~(\ref{aapprox}),   we have
\be
\ba{ll}
\chi f_{n=6}(v_{r})\,\approx\!\!\!&
\frac{36-315 b+1215 b^2-2700 b^3+3780 b^4-3402 b^5+1890 b^6-540 b^7}{\left[b^9+(1-b)^9\right]v_{r}^2}\\
{}&+\,
\frac{-84+756 b-3015 b^2+6975 b^3-10260 b^4+9828 b^5-5922 b^6+1890 b^7}{\left[b^9+(1-b)^9\right]v_{r}^3}\\
{}&+\,\frac{126-1134 b+4536 b^2-10575 b^3+15795 b^4-15552 b^5+9828 b^6-3402 b^7}{\left[b^9+(1-b)^9\right]v_{r}^4}\\
{}&+\,
\frac{-126+1134 b-4536 b^2+10584 b^3-15867 b^4+15795 b^5-10260 b^6+3780 b^7}{\left[b^9+(1-b)^9\right]v_{r}^5}\\
{}&+\,\frac{84-756 b+3024 b^2-7056 b^3+10584 b^4-10575 b^5+6975 b^6-2700 b^7}{\left[b^9+(1-b)^9\right]v_{r}^6} \\
{}&+\,
\frac{-36+324 b-1296 b^2+3024 b^3-4536 b^4+4536 b^5-3015 b^6+1215 b^7}{\left[b^9+(1-b)^9\right]v_{r}^7}\\
{}&+\,
\frac{9-81 b+324 b^2-756 b^3+1134 b^4-1134 b^5+756 b^6-315 b^7}{\left[b^9+(1-b)^9\right]v_{r}^8}\\
{}&+\,
\frac{-1+9 b-36 b^2+84 b^3-126 b^4+126 b^5-84 b^6+36 b^7}{\left[b^9+(1-b)^9\right]v_{r}^9}\,.
\ea
\ee

Letting  $\chi=3.2991$,   we get an ${n=6}$ Janus van der Waals equation for helium-$4$,
\be
\ba{l}
\left({P_{r}-
	\frac{10.671}{v_{r}^{2}}
	+\frac{97.188}{v_{r}^{3}}
	-\frac{259.53}{v_{r}^{4}}+
	\frac{393.69}{v_{r}^{5}}-
	\frac{366.57}{v_{r}^{6}}+
	\frac{210.75}{v_{r}^{7}}-
	\frac{69.112}{v_{r}^{8}}+
	\frac{10.066}{v_{r}^{9}}}
\right)
\!\left(v_{r}-0.51519\right)\approx 3.2991\, T_{r}\,.
\ea
\label{n6JvdW}
\ee 

%%%%%%%%%%%%%%%%%%%%%%%%
\section{Comparison with NIST Reference Data}

Henceforth we  look into isochoric, isobaric and isothermal cases for real molecules, which will demonstrate that our Janus van der Waals equations represent excellent agreements with the NIST data,   better than the original van der Waals equation.  By construction,  the  Janus van der Waals equations reflect  the previously reported two-sided  critical phenomena~\cite{Cho:2016jzz} and at the same time reduce consistently to the classical ideal gas law in the low density limit  far away from the critical point at ${v_{r}=1}$.

First, we focus on the $n=2$ case  which only cyclopentane molecule ($\mathrm{C_{5} H_{10}}$) belongs to.

\begin{figure}[H]
	\begin{center}
		\includegraphics[width=8.80cm]{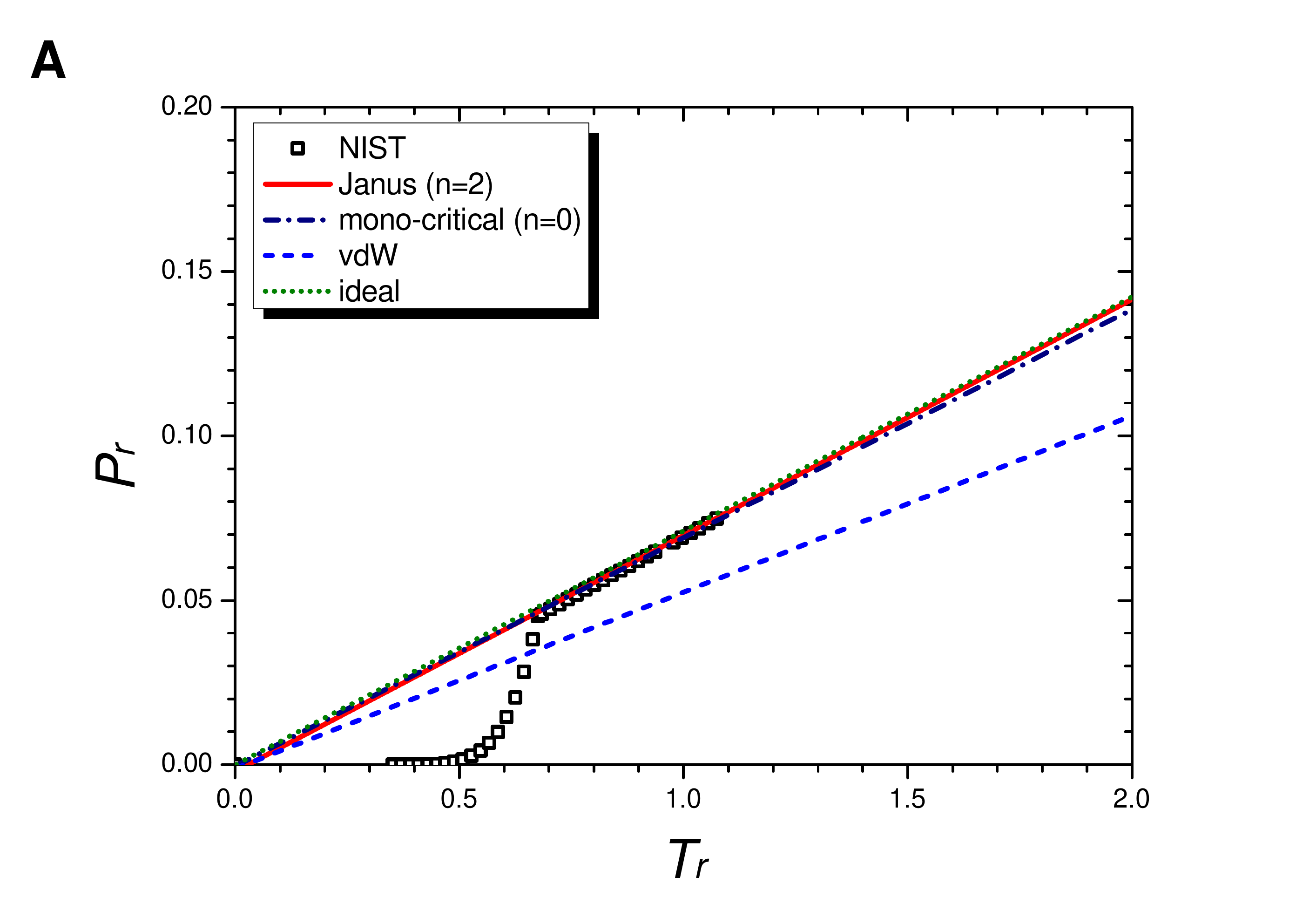}
		\includegraphics[width=8.80cm]{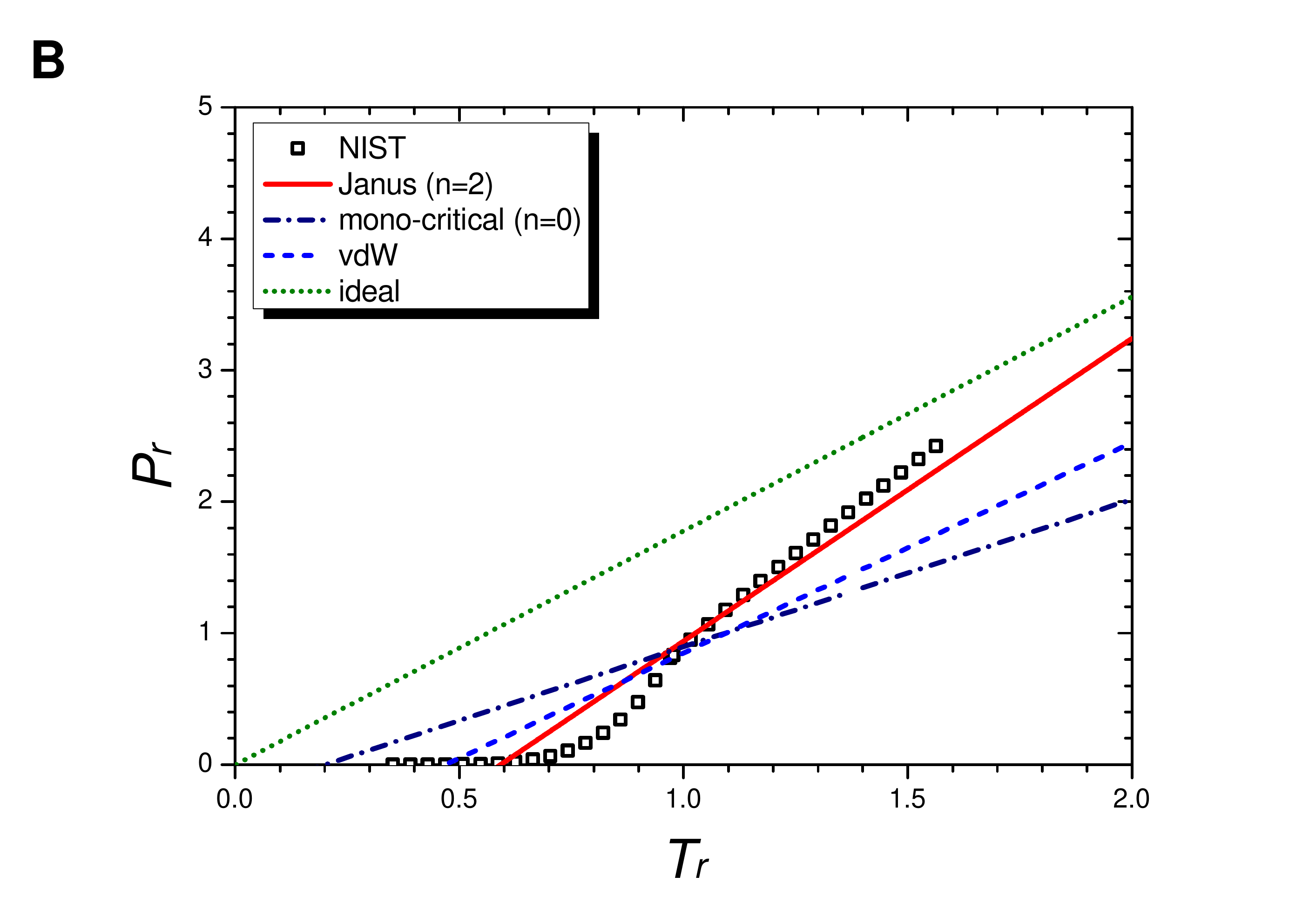}
		\vspace{1cm}
		
	     \includegraphics[width=8.80cm]{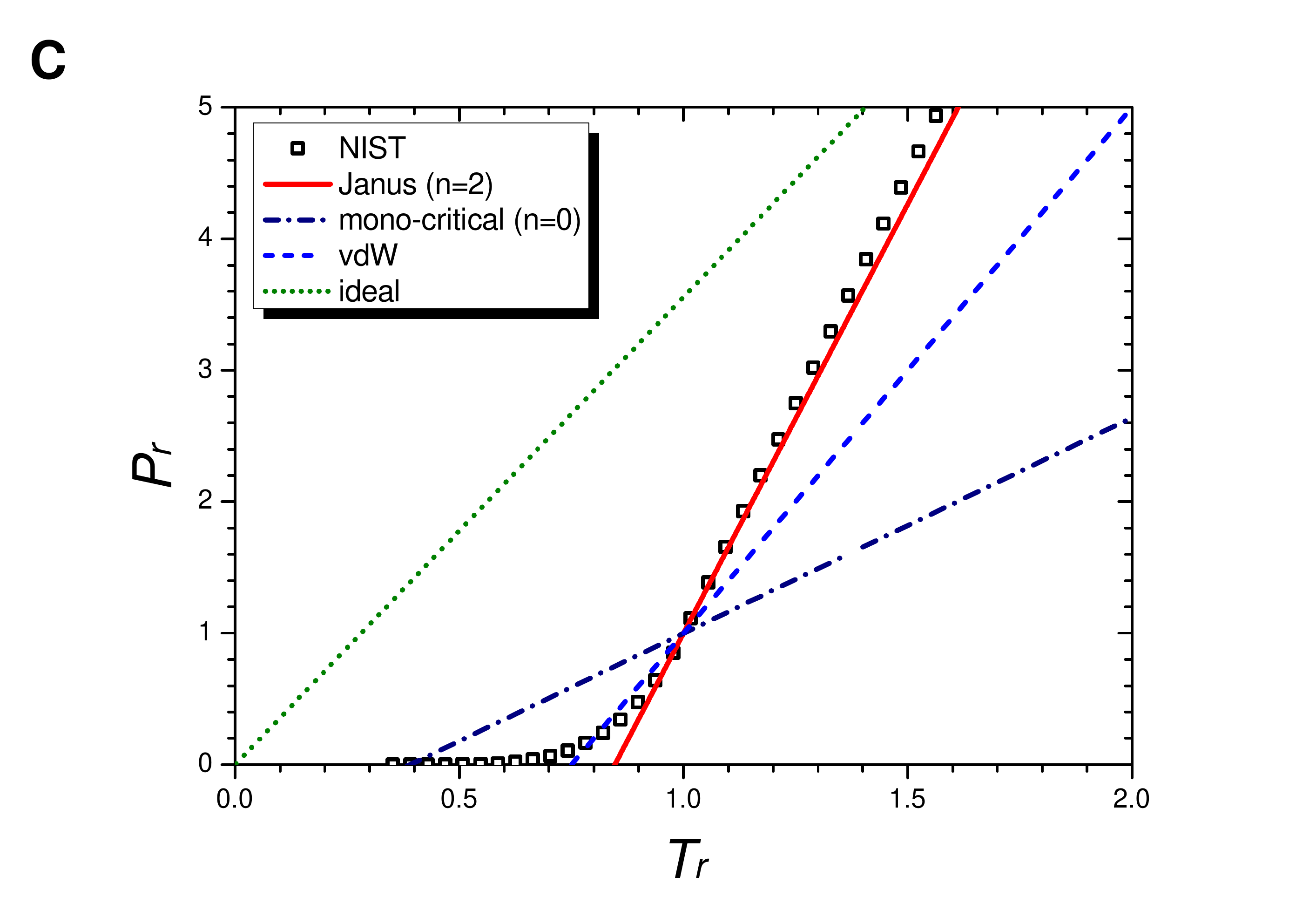}
		\includegraphics[width=8.80cm]{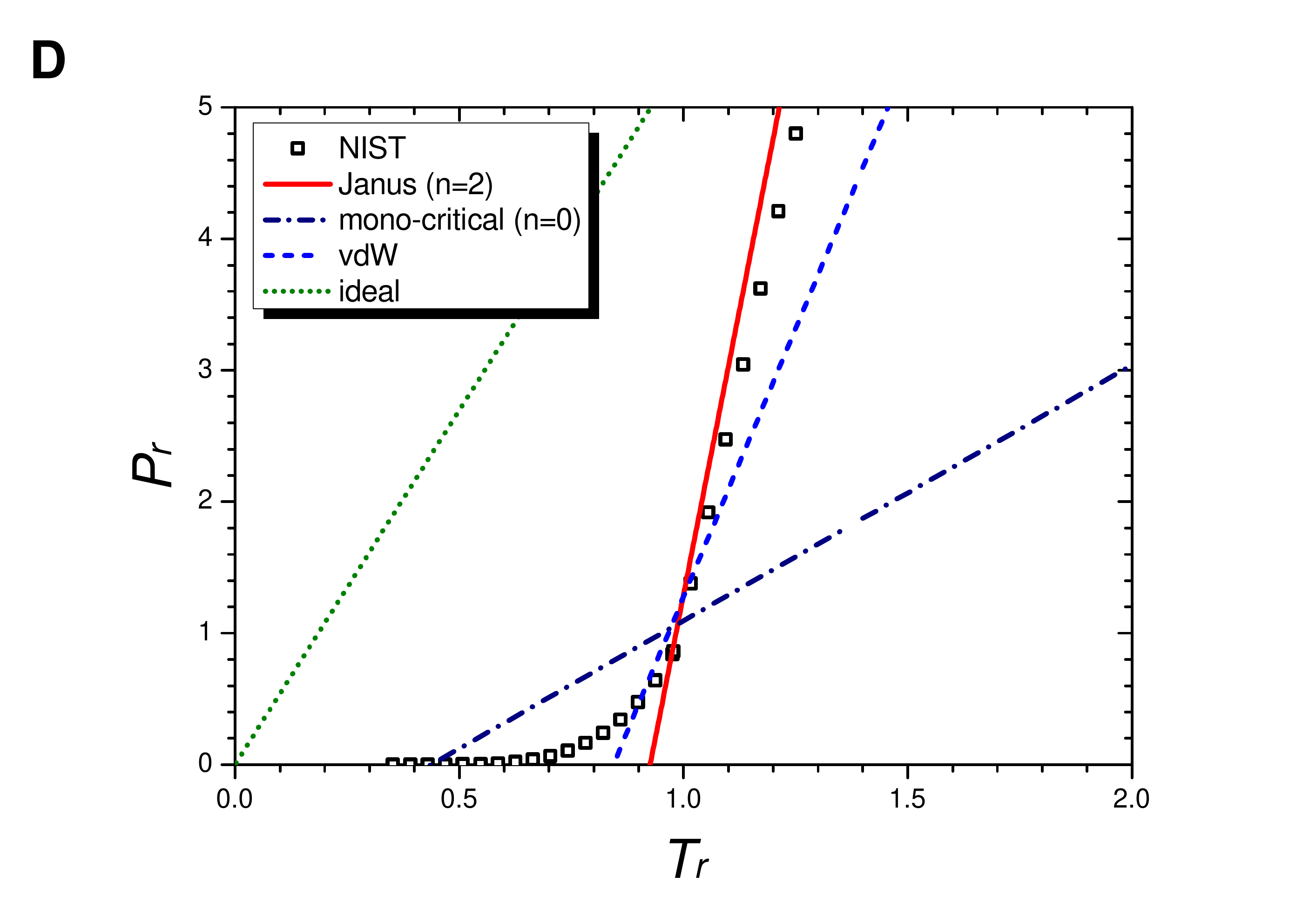}
		\caption{Isochoric curves of  cyclopentane~($\mathrm{C_{5} H_{10}}$)  at $1/v_{r} = 0.02$ ({\bf{A}}), $1/v_{r} = 0.5$ ({\bf{B}}),  $1/v_{r} = 1.0$  ({\bf{C}}), and  $1/v_{r} = 1.5$  ({\bf{D}}). Boxes are from the NIST  data. The red solid line is drawn  from the $n=2$ Janus van der Waals equation~(\ref{n2JvdW}) and is better fitted than  the three others: the $n=0$ mono-critical equation~(\ref{n0JvdW}),  the original van der Waals equation (\ref{vdW}), and the classical ideal gas law~(\ref{ideal}).}\label{cyclopentane_isochore}
	\end{center}
\end{figure}

\textbf{Figure~\ref{cyclopentane_isochore}} shows the isochoric curves of cyclopentane molecule  at $1/v_{r} = 0.02$ ({\bf{A}}), $1/v_{r} = 0.5$ ({\bf{B}}),  $1/v_{r} = 1.0$  ({\bf{C}}), and  $1/v_{r} = 1.5$  ({\bf{D}})  respectively. They are  drawn by the NIST data  and further by the four equations: the $n=2$ Janus van der Waals equation~(\ref{n2JvdW}), the mono-critical equation ($n=0$)~(\ref{n0JvdW}), the original van der Waals equation (\ref{vdW}), and the classical ideal gas law (\ref{ideal}). The $n=2$ Janus van der Waals equation fits best  with the NIST data, while  consistently  reducing   to the classical ideal gas law in low density limit. \\

\newpage
%%%

\begin{figure}[H]
	\begin{center}
		\includegraphics[width=8.8cm]{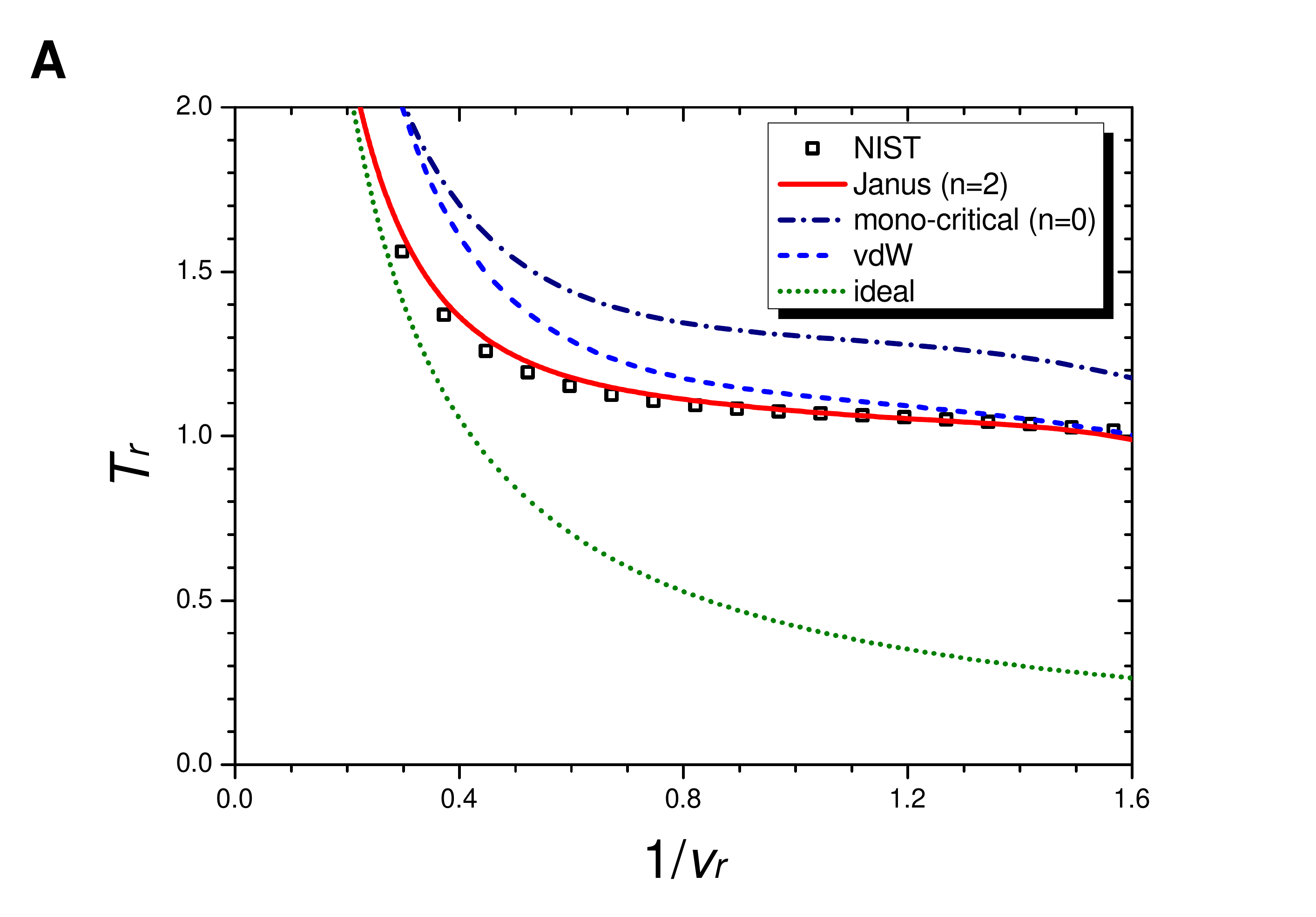}
		\includegraphics[width=8.8cm]{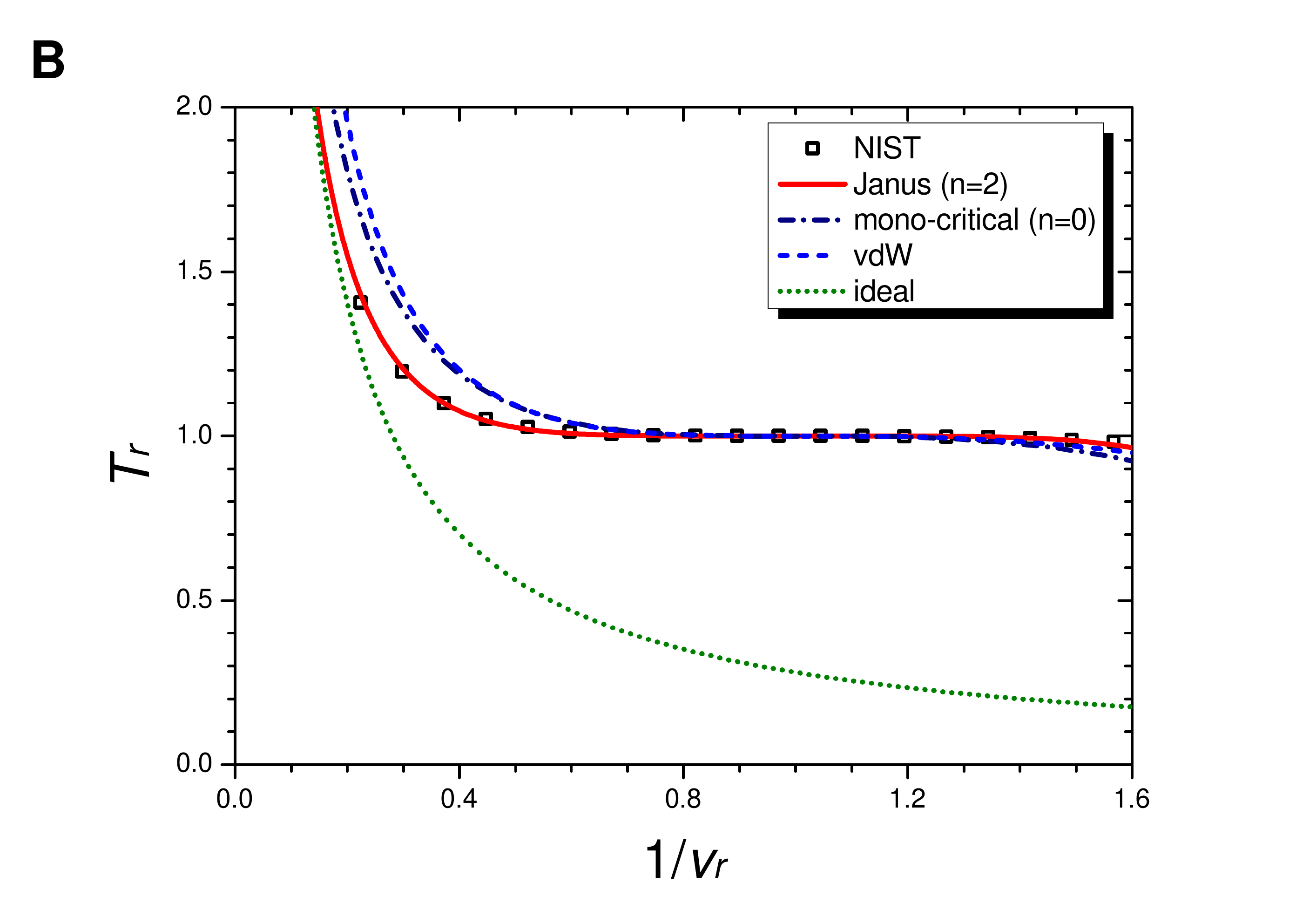}
		\includegraphics[width=8.8cm]{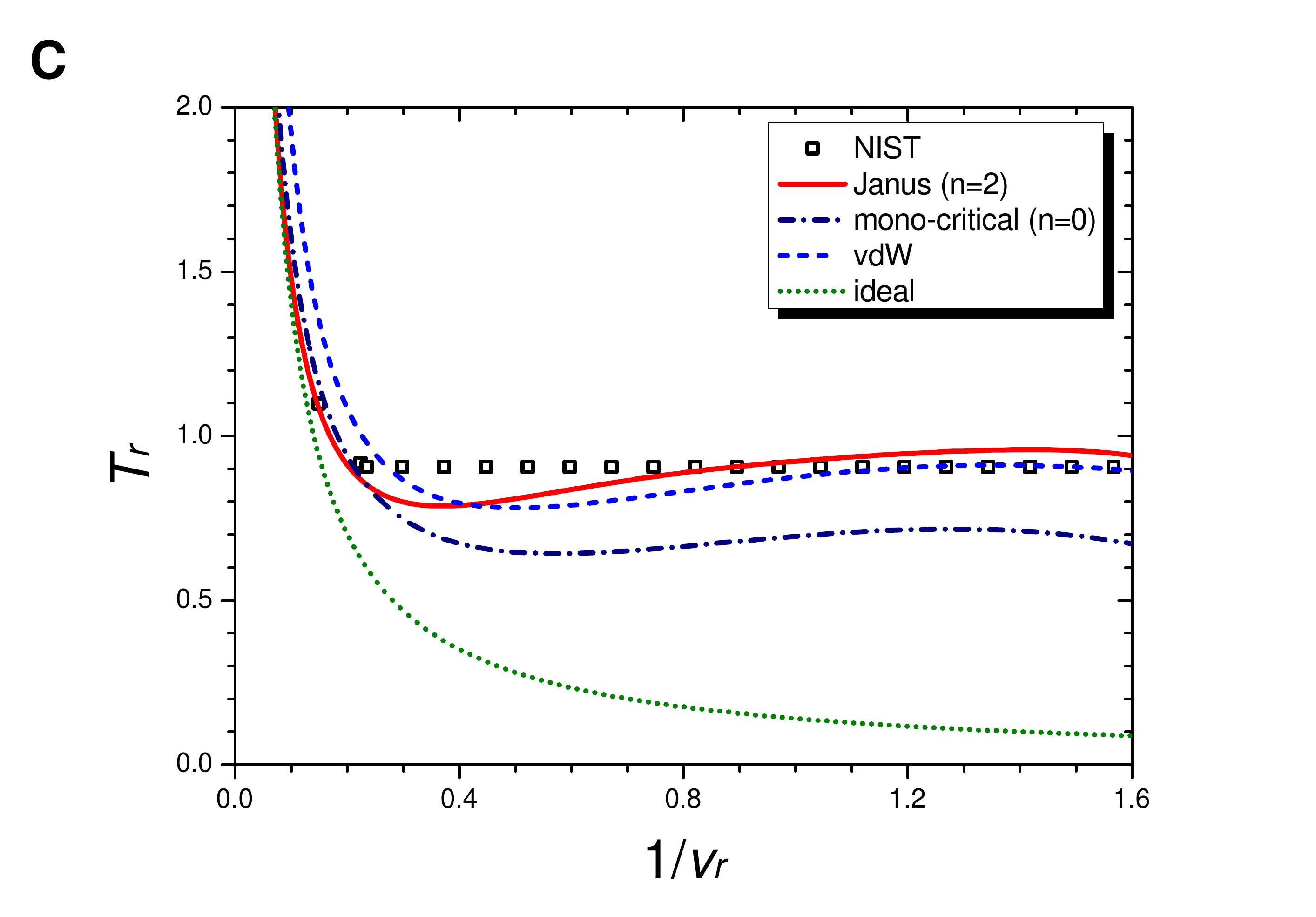}
		\caption{Isobaric curves  of cyclopentane~($\mathrm{C_{5} H_{10}}$)  at  $P_{r} = 1.5$  ({\bf{A}}), $P_{r} = 1.0$  ({\bf{B}}), and  $P_{r} = 0.5$ ({\bf{C}}). Boxes are from the NIST  data. The red solid line is drawn  from the $n=2$ Janus van der Waals equation~(\ref{n2JvdW}) and is better fitted than  the $n=0$ mono-critical equation~(\ref{n0JvdW}),  the original van der Waals equation (\ref{vdW}), or the classical ideal gas law~(\ref{ideal}).}\label{cyclopentane_isobar}
	\end{center}
\end{figure}

\textbf{Figure~\ref{cyclopentane_isobar}} shows the  isobaric curves  of  cyclopentane molecule ($\mathrm{C_{5} H_{10}}$) at $P_{r} = 1.5$  ({\bf{A}}), $P_{r} = 1.0$ ({\bf{B}}), and  $P_{r} = 0.5$  ({\bf{C}})  respectively. They are 
drawn by the NIST data and further by the four equations: the $n=2$ Janus van der Waals equation~(\ref{n2JvdW}), the mono-critical equation ($n=0$)~(\ref{n0JvdW}), the original van der Waals equation (\ref{vdW}), and the classical ideal gas law (\ref{ideal}). The $n=2$ Janus van der Waals equation is in  excellent agreement with the NIST data, especially at the liquid-vapor coexistence region near $P_{r}=1$ as well as at the supercritical region of $P_{r} > 1$. %The Janus van der Waals equation has a strong point to depict realistic isobars near $P=P_{c}$ and at $P>P_{c}$.  

%%%
\newpage

\begin{figure}[H]
	\begin{center}
		\includegraphics[width=8.80cm]{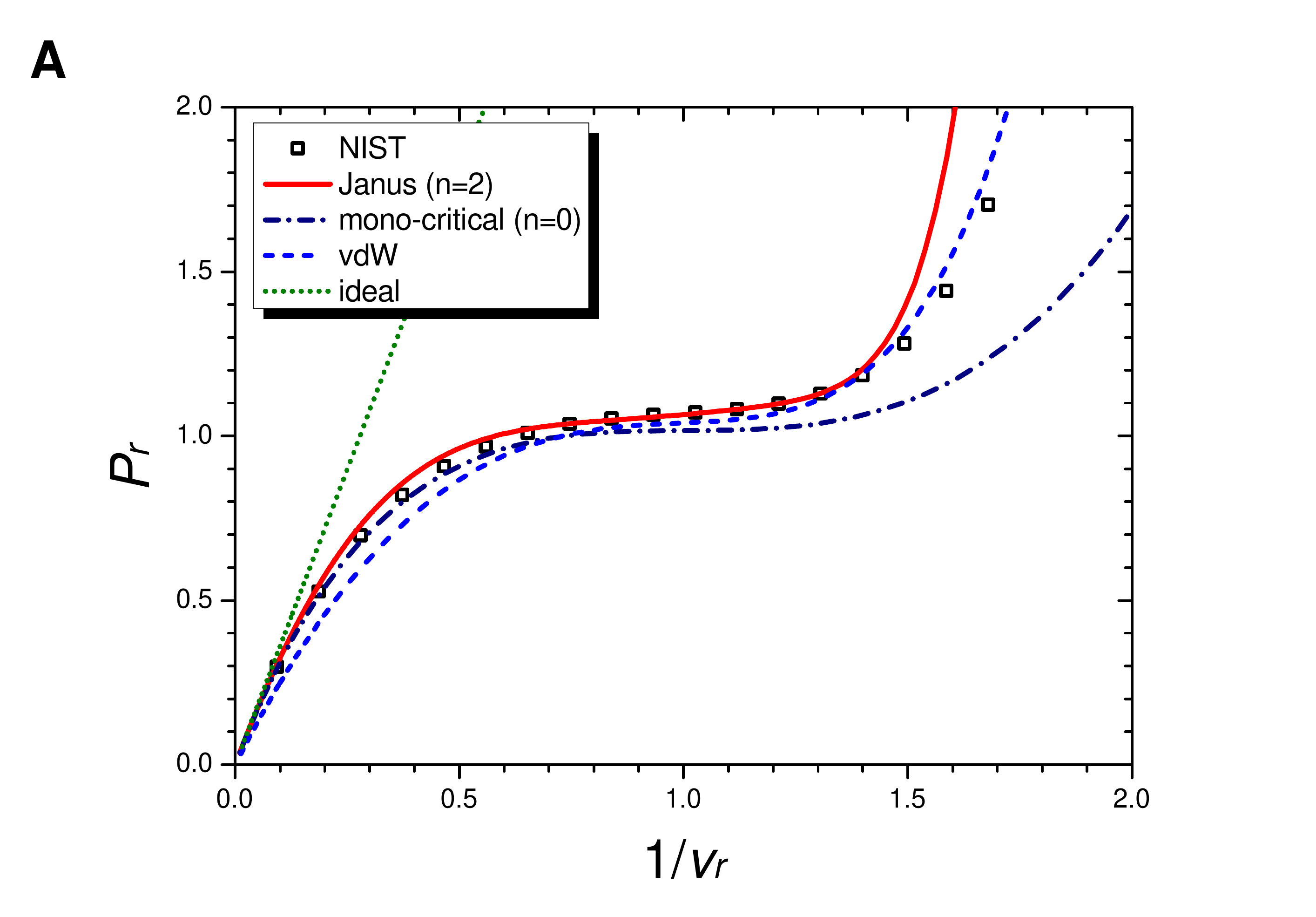}
		\includegraphics[width=8.80cm]{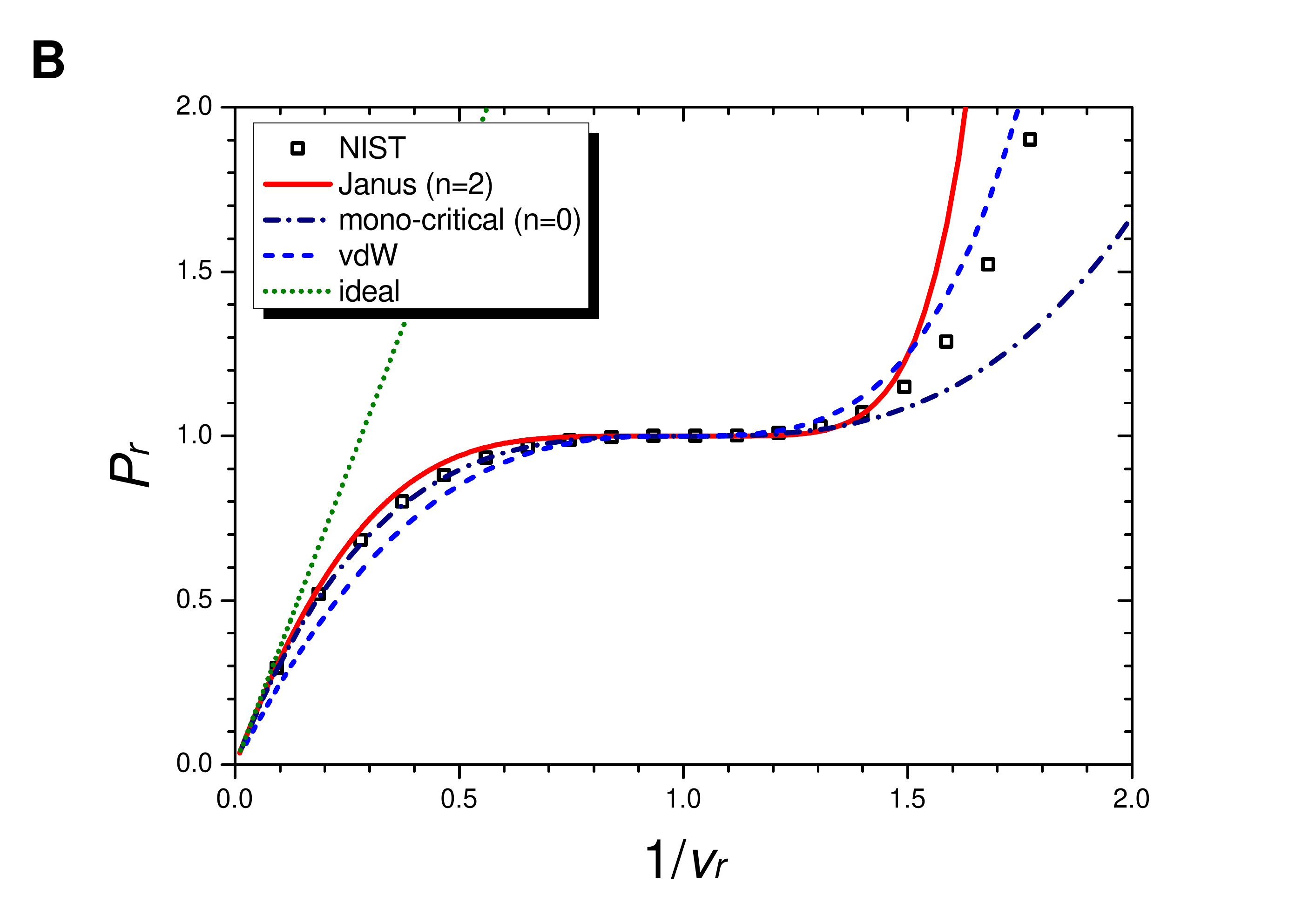}
		\includegraphics[width=8.80cm]{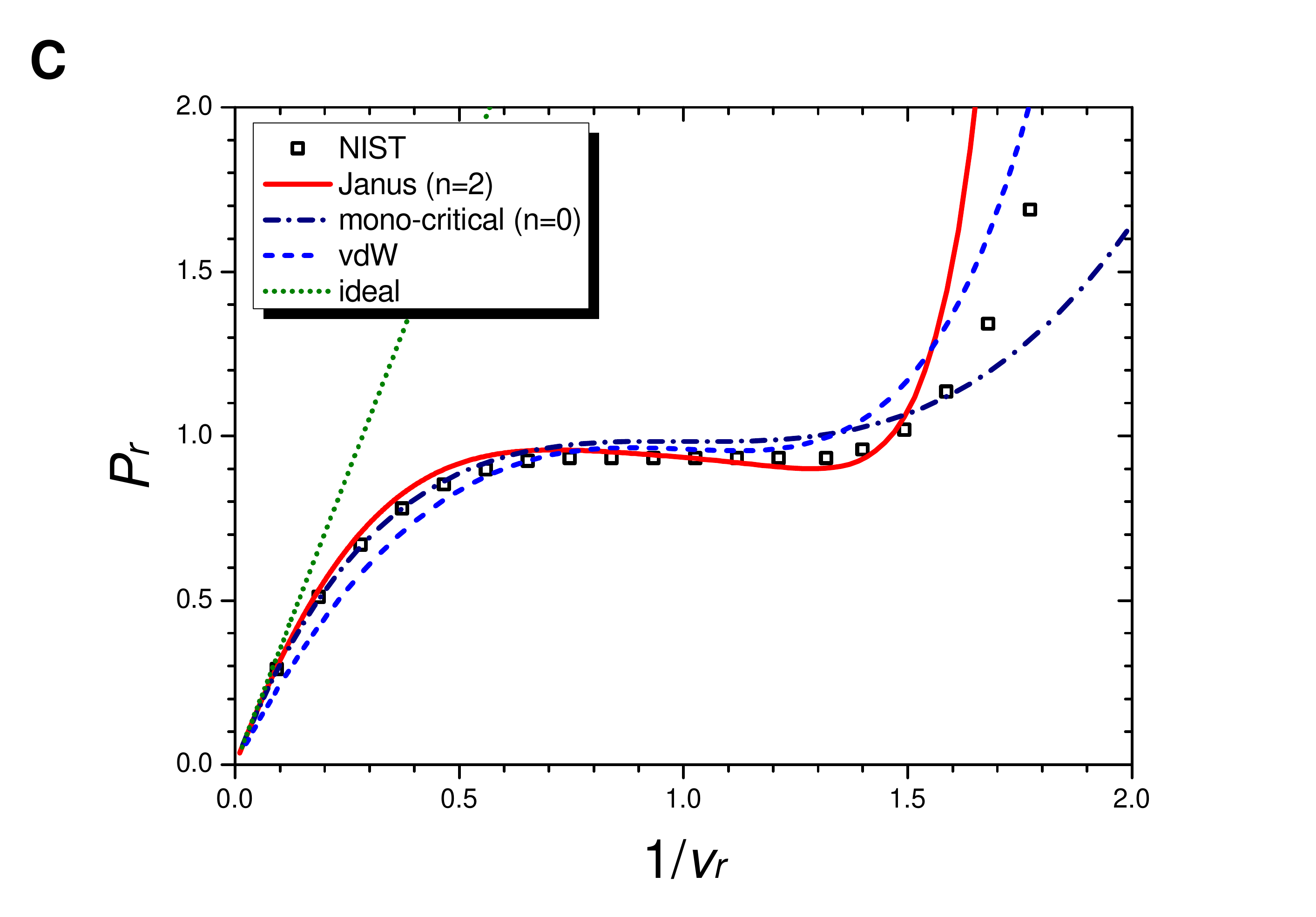}
		\caption{Isothermal  curves of cyclopentane~($\mathrm{C_{5} H_{10}}$)  at  $T_{r} = 1.01$  ({\bf{A}}), $T_{r} = 1.00$  ({\bf{B}}), and  $T_{r} = 0.99$ ({\bf{C}}). Boxes are from the NIST  data. The red solid line is drawn  from the $n=2$ Janus van der Waals equation~(\ref{n2JvdW}) and is better fitted than  the $n=0$ mono-critical equation~(\ref{n0JvdW}),  the original van der Waals equation (\ref{vdW}), or the classical ideal gas law~(\ref{ideal}).	}\label{cyclopentane_isotherm}
	\end{center}
\end{figure}

\textbf{Figure~\ref{cyclopentane_isotherm}} shows the  isothermal  curves  of  cyclopentane molecule ($\mathrm{C_{5} H_{10}}$) at $T_{r} = 1.01$  ({\bf{A}}), $T_{r} = 1.00$ ({\bf{B}}), and  $T_{r} = 0.99$  ({\bf{C}})  respectively. They are 
drawn by the NIST data and further by the four equations: the $n=2$ Janus van der Waals equation~(\ref{n2JvdW}), the mono-critical equation ($n=0$)~(\ref{n0JvdW}), the original van der Waals equation (\ref{vdW}), and the classical ideal gas law (\ref{ideal}).  The Janus van der Waals equation shows enhanced sigmoid shape compared to the original van der Waals equation when $T_{r}$ is lower than $1$. \\%When $T_{r} \ge 1$,  the Janus van der Waals equation fits very well with the NIST data near the  critical point. \\

\vspace{-17pt}
%\vspace{-31pt}

\newpage
%%%

\begin{figure}[H]
	\begin{center}
		\includegraphics[width=18.9cm]{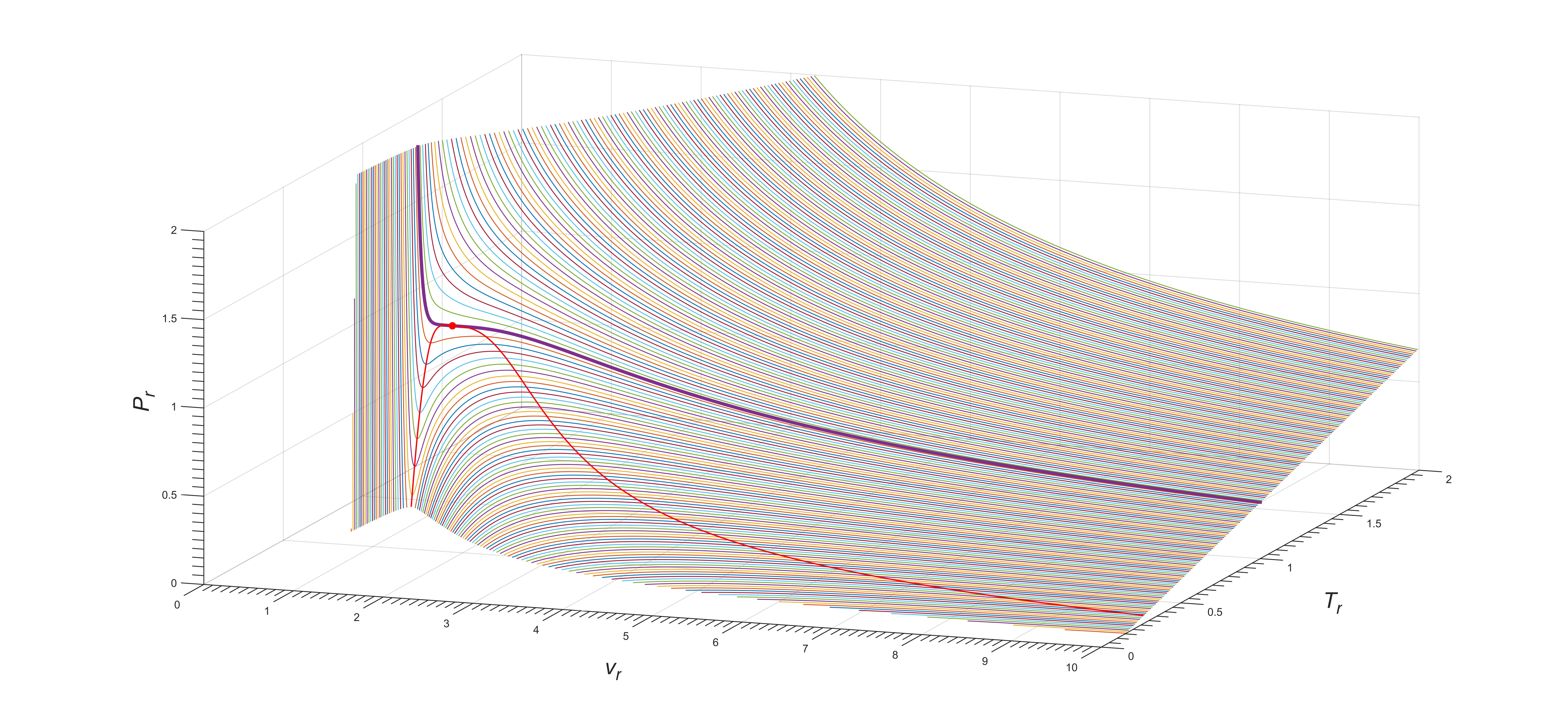}
		\caption{Three-dimensional $P_{r}-v_{r}-T_{r}$ phase diagram of the  exact $n=2$ Janus van der Waals equation~(\ref{bcvdW}), as for  cyclopentane molecule ($\mathrm{C_{5} H_{10}}$).  The bold purple line corresponds to the isotherm of $T_{r} = 1.00$ as depicted in  \textbf{Figure~\ref{cyclopentane_isotherm}~{\bf{B}}}; the red line is the Janus van der Waals spinodal curve with $a = 0.99$;  and the red dot is the critical point.}\label{cyclopentane_PVT_spinodal}
	\end{center}
\end{figure}

\textbf{Figure~\ref{cyclopentane_PVT_spinodal}} shows  the three-dimensional $P_{r}-v_{r}-T_{r}$ phase diagram of the exact  ${n=2}$ Janus van der Waals equation~(\ref{bcvdW}) with (\ref{fform}), as for  cyclopentane molecule ($\mathrm{C_{5} H_{10}}$).  The red line shows  the spinodal curve~(\ref{spinodalT}), \textit{i.e.~}$\partial_{v}P(T,v)=0$,  of the Janus van der Waals equation with the choice of $a = 0.99$. 

\newpage

%%%%

Now, we turn  to the case of  ${n=4}$  which the following nine molecules belong to: nitrogen~($\mathrm{N}_{2}$), argon~($\mathrm{Ar}$), methane~($\mathrm{CH}_{4}$), ethylene~($\mathrm{C}_{2}\mathrm{H}_{4}$), ethane~($\mathrm{C}_{2}\mathrm{H}_{6}$),  propylene~($\mathrm{C}_{3}\mathrm{H}_{6}$), propane~($\mathrm{C}_{3}\mathrm{H}_{8}$), butane~($\mathrm{C}_{4}\mathrm{H}_{10}$),  and isobutane~($\mathrm{C}_{4}\mathrm{H}_{10}$). Here we choose nitrogen as a representative example. The other eight molecules as  well as the $n=6$ case which only helium-4 molecule belongs to  are  dealt in the Supplementary Material separately. \\

\begin{figure}[H]
	\begin{center}
		\includegraphics[width=8.80cm]{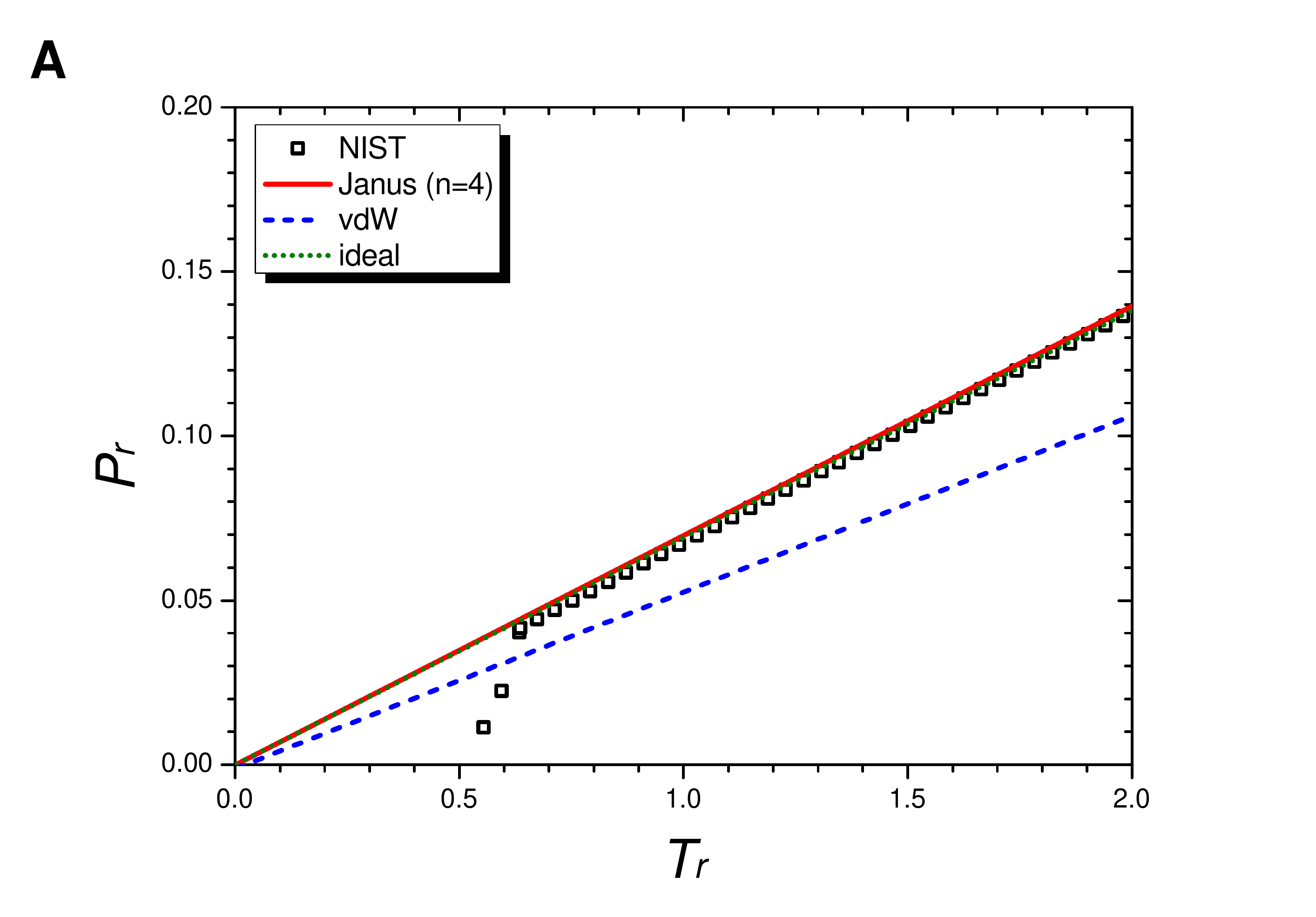}
		\includegraphics[width=8.80cm]{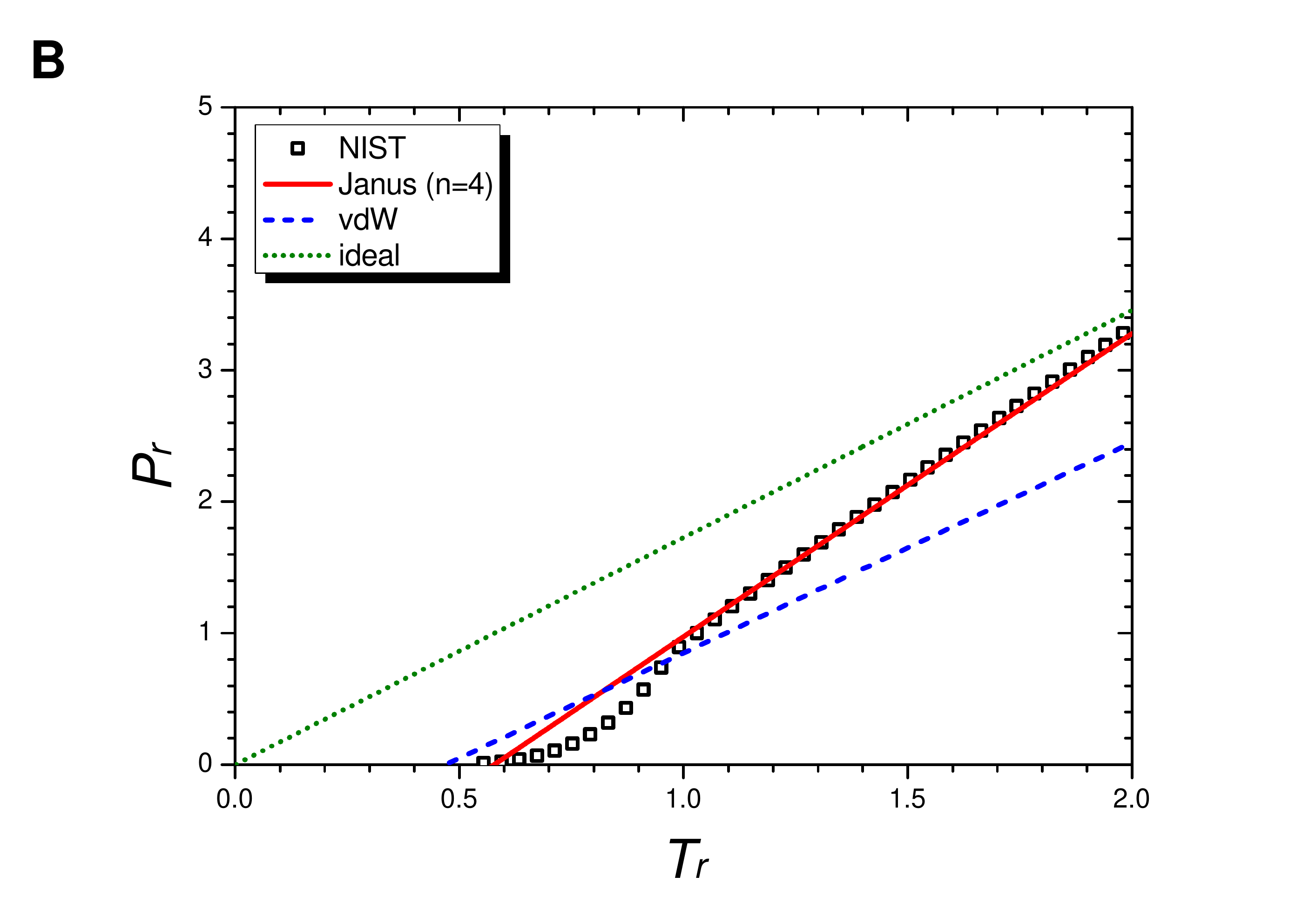}
		\includegraphics[width=8.80cm]{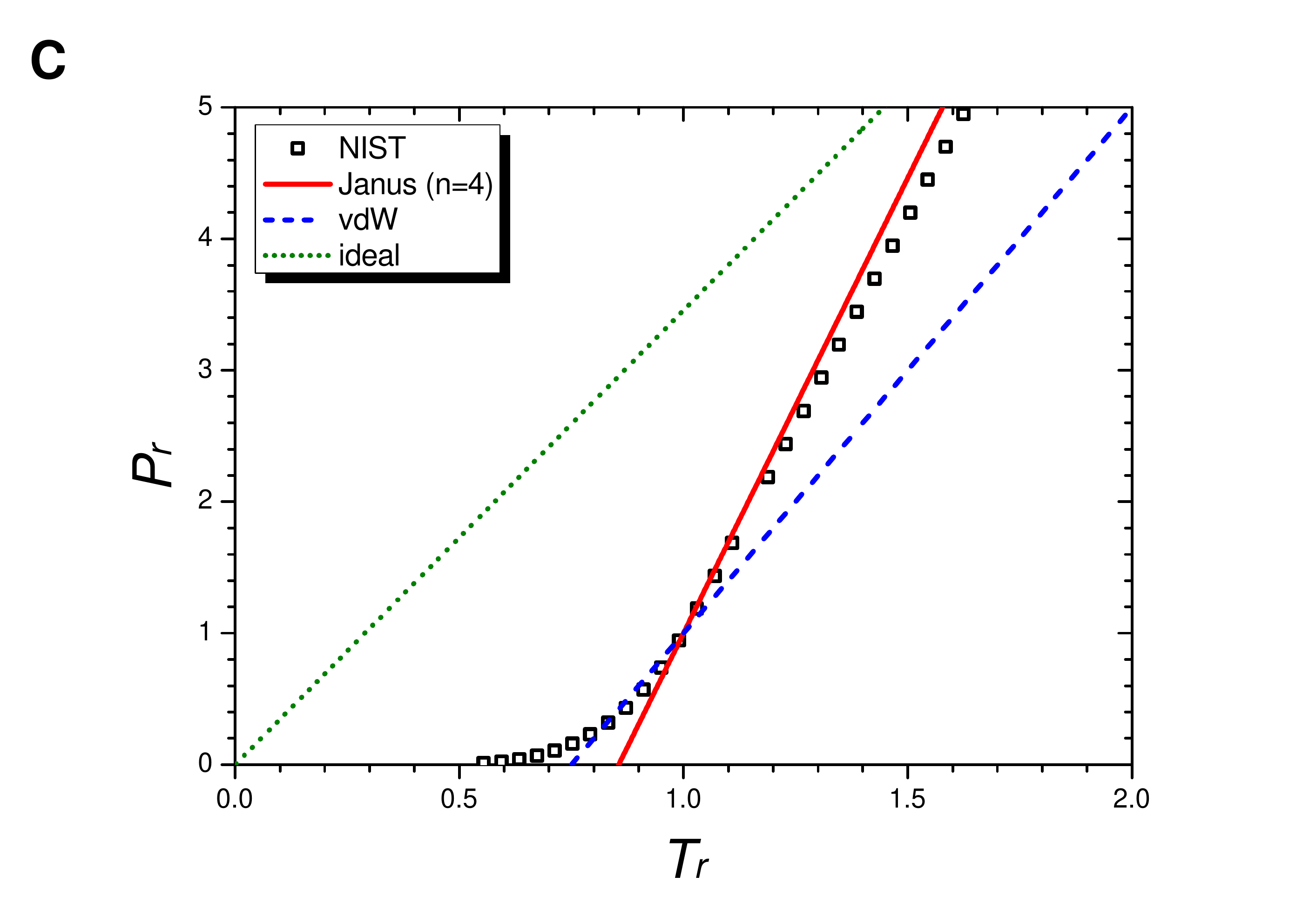}
		\includegraphics[width=8.80cm]{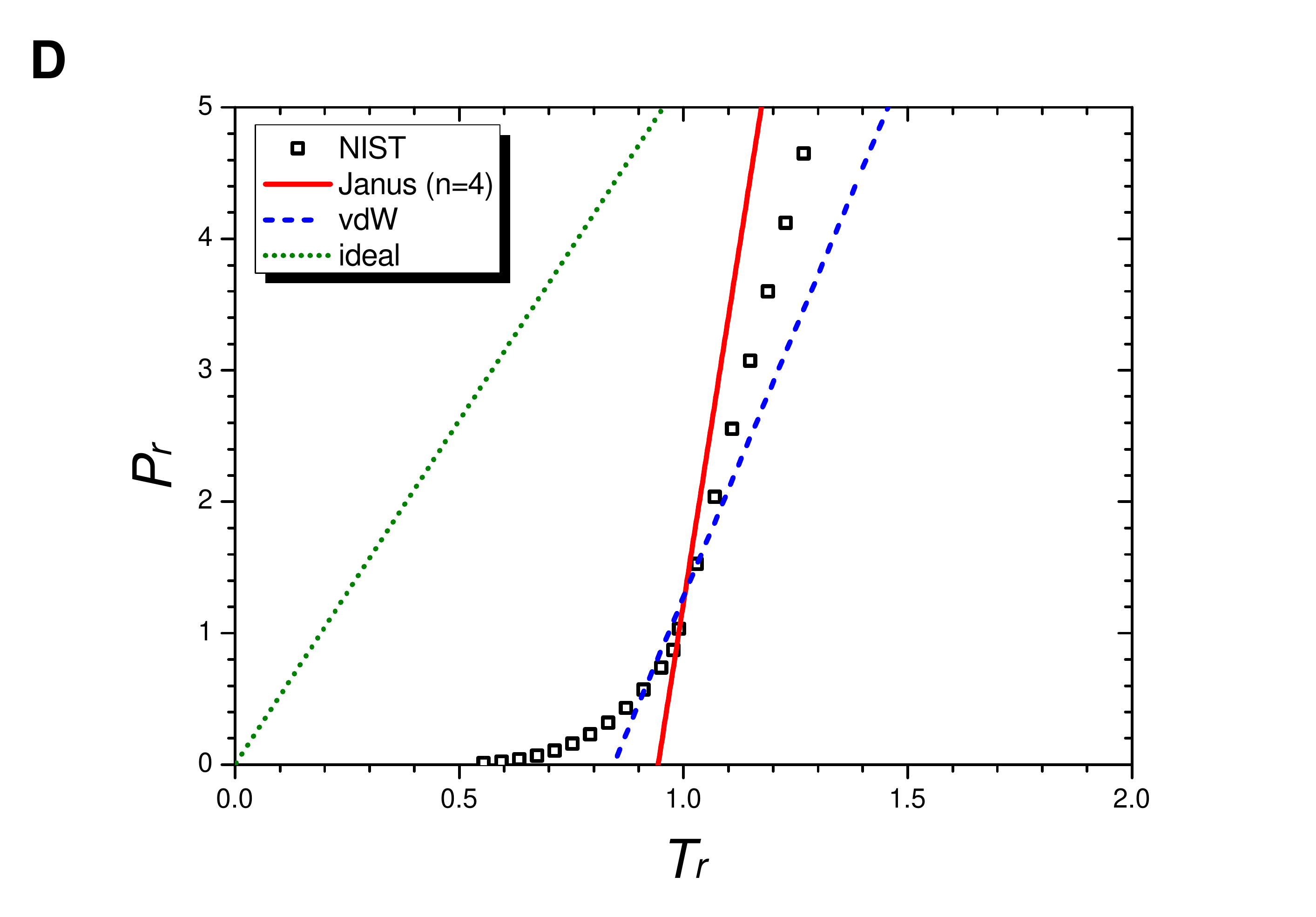}
		\caption{Isochoric curves of nitrogen~($\mathrm{N}_{2}$) at $1/v_{r} = 0.02$ ({\bf{A}}), $1/v_{r} = 0.5$ ({\bf{B}}),  $1/v_{r} = 1.0$  ({\bf{C}}), and  $1/v_{r} = 1.5$  ({\bf{D}}). Boxes are from the NIST  data. The red solid line is drawn  from the $n=2$ Janus van der Waals equation~(\ref{n2JvdW}) and is better fitted than   the original van der Waals equation (\ref{vdW}) or the classical ideal gas law~(\ref{ideal}).}\label{nitrogen_isochore}
	\end{center}
\end{figure}

\textbf{Figure~\ref{nitrogen_isochore}}   shows the isochoric curves of nitrogen~($\mathrm{N}_{2}$)  at $1/v_{r} = 0.02$ ({\bf{A}}), $1/v_{r} = 0.5$ ({\bf{B}}),  $1/v_{r} = 1.0$  ({\bf{C}}), and  $1/v_{r} = 1.5$  ({\bf{D}})  respectively. They are  drawn by the NIST data  and further by the four equations: the Janus van der Waals equation for ${n=4}$~(\ref{n4JvdW}), the original van der Waals equation (\ref{vdW}), and the classical ideal gas law (\ref{ideal}).  The Janus van der Waals equation fits best with the NIST data, especially at the region of $T_{r} > 1$ better than the original van der Waals equation.  Again, we confirm that the $n=4$ Janus van der Waals equation reduces  to the classical ideal gas law in low density limit.

\newpage
%%%

\begin{figure}[H]
	\begin{center}
		\includegraphics[width=8.80cm]{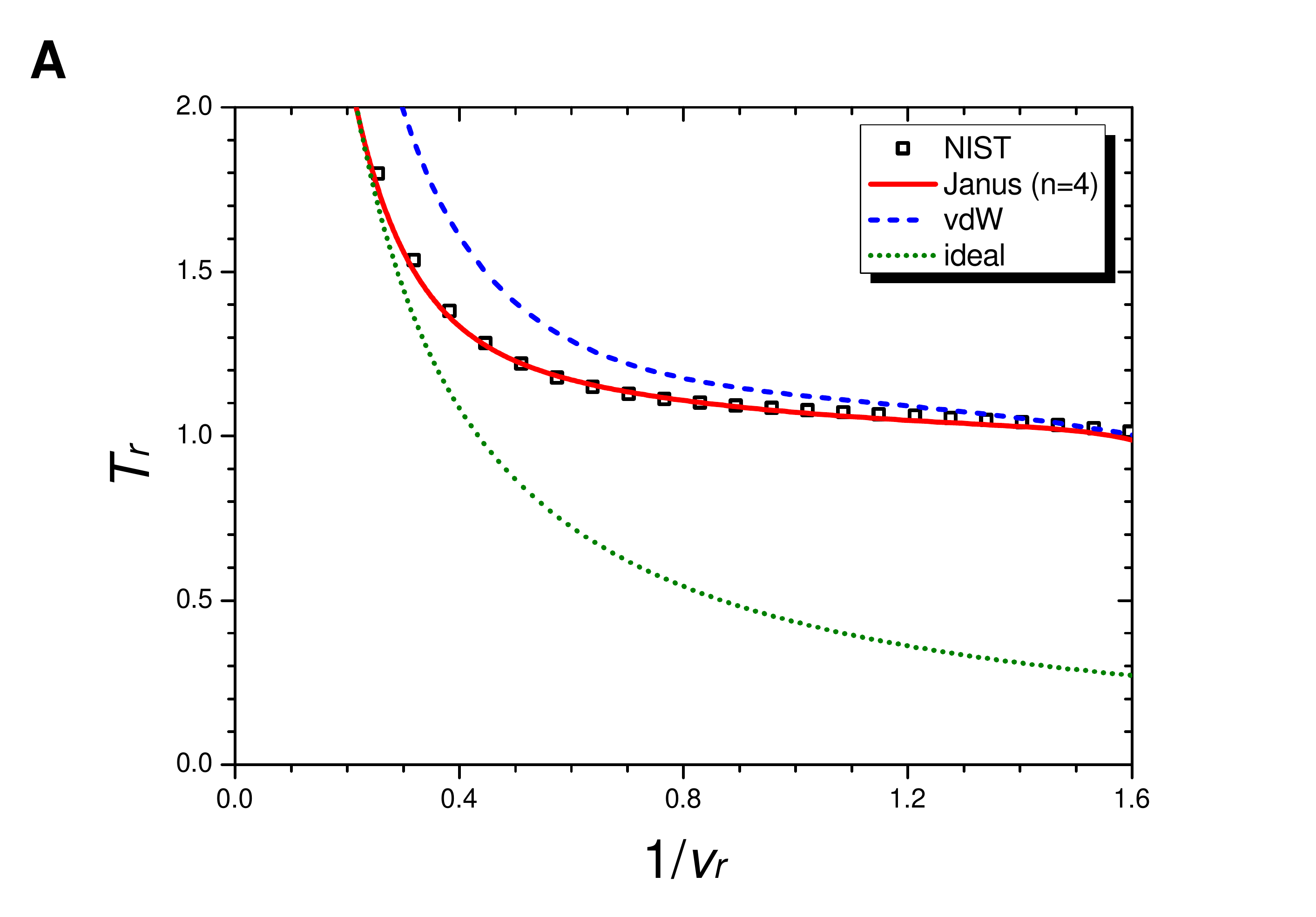}
		\includegraphics[width=8.80cm]{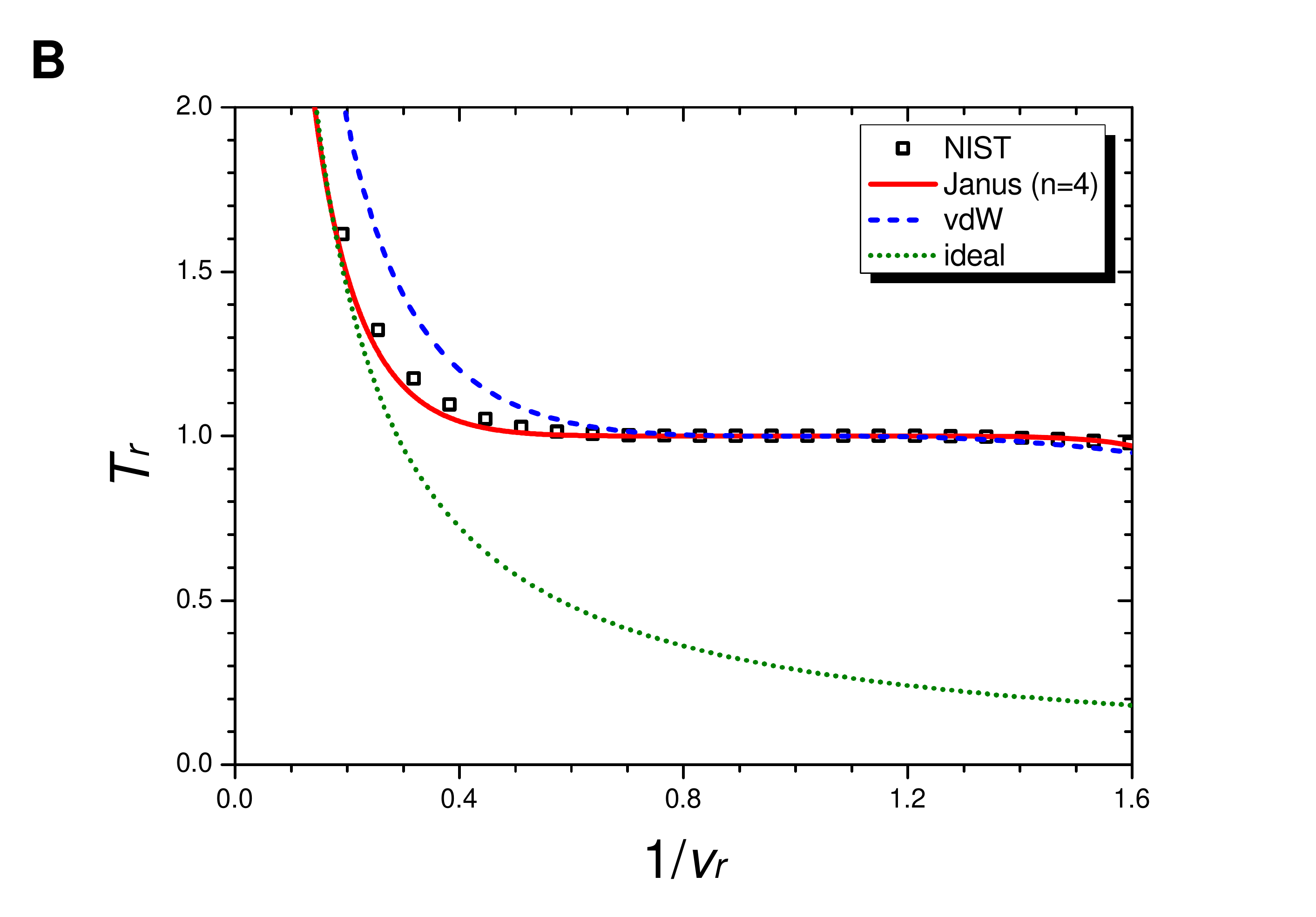}
		\includegraphics[width=8.80cm]{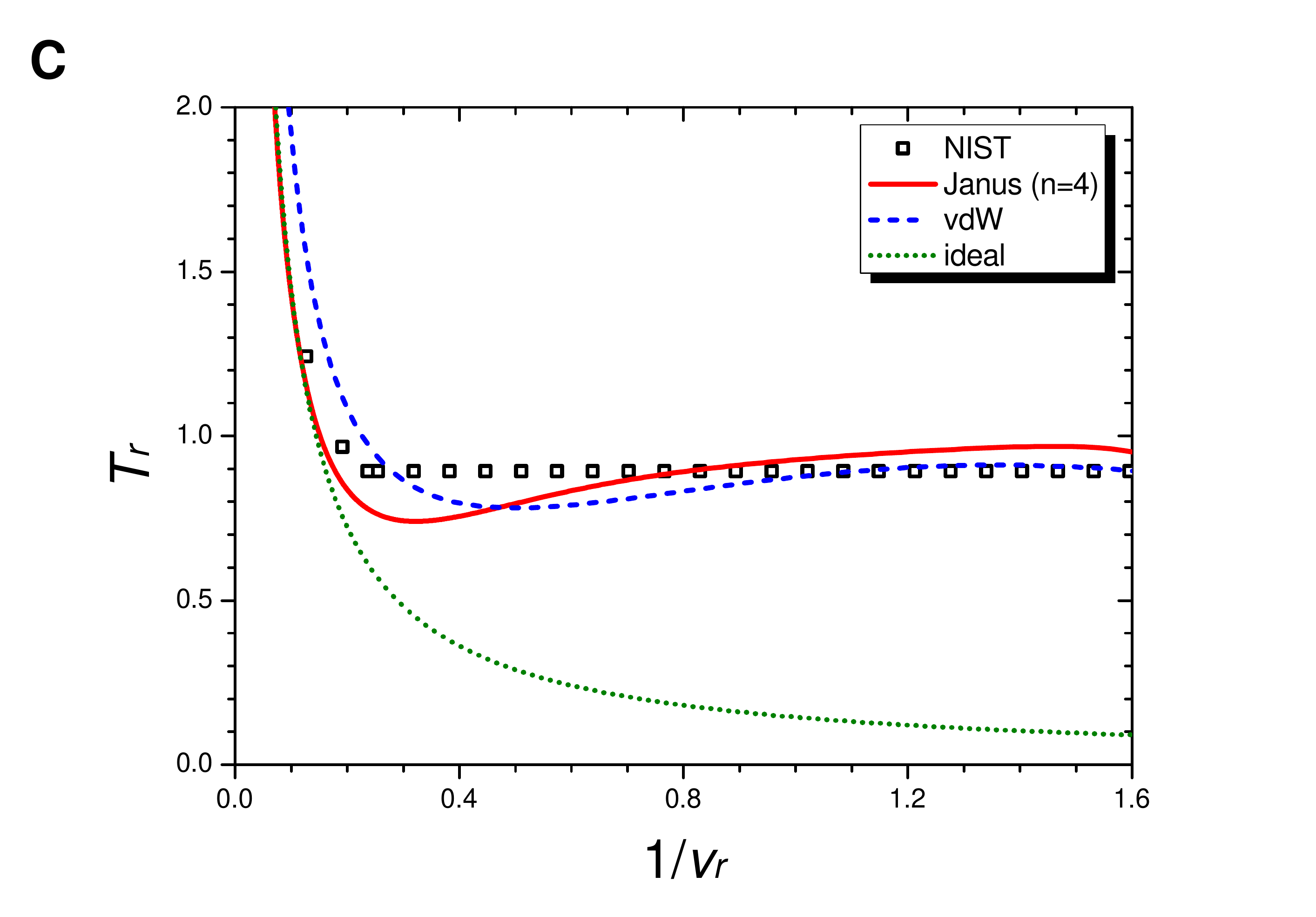}
		\caption{Isobaric curves of   nitrogen~($\mathrm{N}_{2}$)  at  $P_{r} = 1.5$  ({\bf{A}}), $P_{r} = 1.0$  ({\bf{B}}), and  $P_{r} = 0.5$ ({\bf{C}}). Boxes are from the NIST  data. The red solid line is drawn  from the $n=4$ Janus van der Waals equation~(\ref{n4JvdW}) and is better fitted than  the original van der Waals equation (\ref{vdW}) or the classical ideal gas law~(\ref{ideal}).}\label{nitrogen_isobar}
	\end{center}
\end{figure}

\textbf{Figure~\ref{nitrogen_isobar}} shows the  isobaric curves  of  nitrogen ($\mathrm{N}_{2}$)  at $P_{r} = 1.5$  ({\bf{A}}), $P_{r} = 1.0$ ({\bf{B}}), and  $P_{r} = 0.5$  ({\bf{C}})  respectively. They are 
drawn by the NIST data and further by the three equations: the Janus van der Waals equation for $n=4$~(\ref{n4JvdW}),  the original van der Waals equation (\ref{vdW}), and the classical ideal gas law (\ref{ideal}). The $n=4$ Janus van der Waals equation is in  excellent agreement with the NIST data, especially at the liquid-vapor coexistence region near $P_{r}=1$ as well as at the supercritical region of $P_{r} > 1$.\\

\newpage
%%%

\begin{figure}[H]
	\begin{center}
		\includegraphics[width=8.80cm]{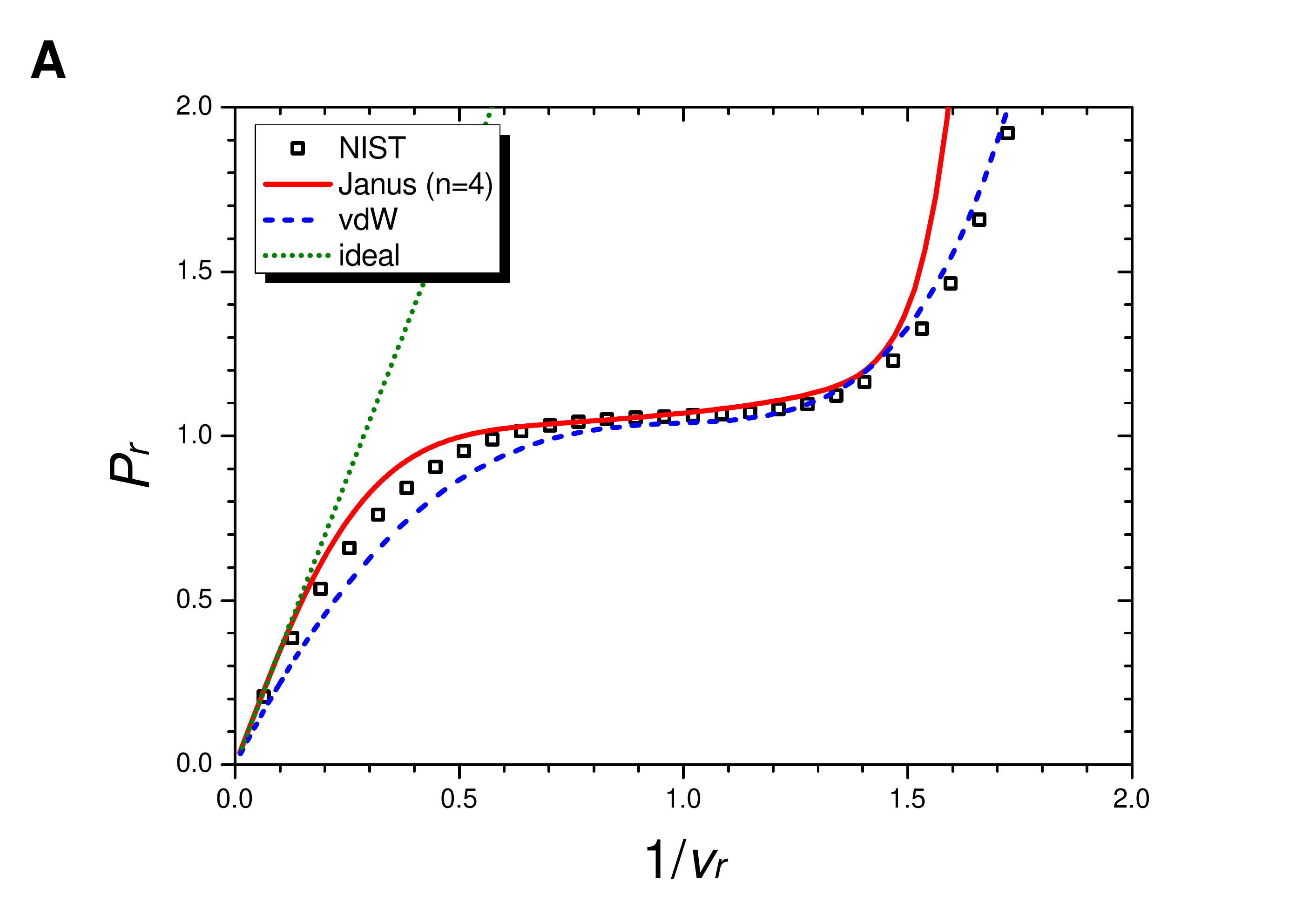}
		\includegraphics[width=8.80cm]{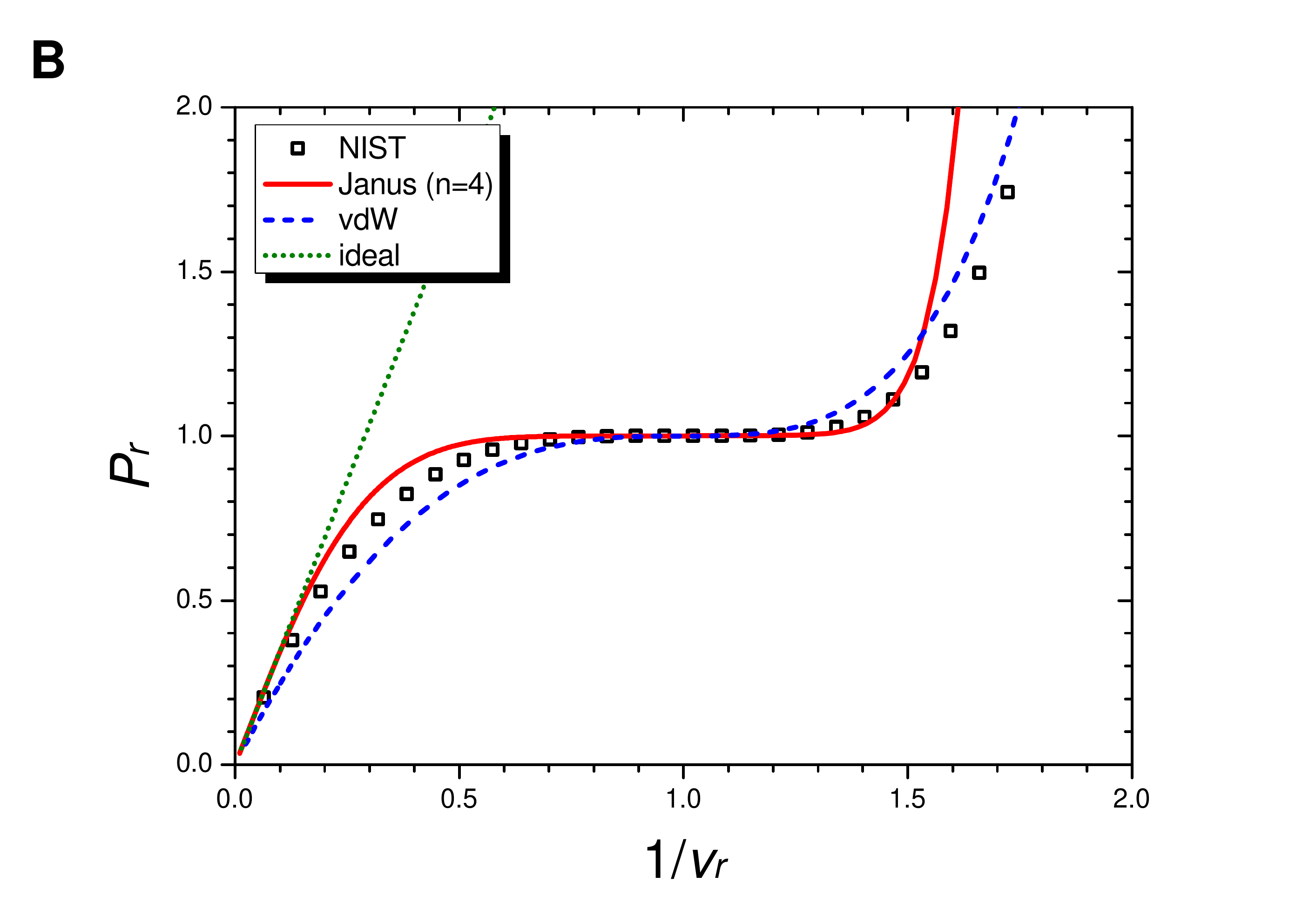}
		\includegraphics[width=8.80cm]{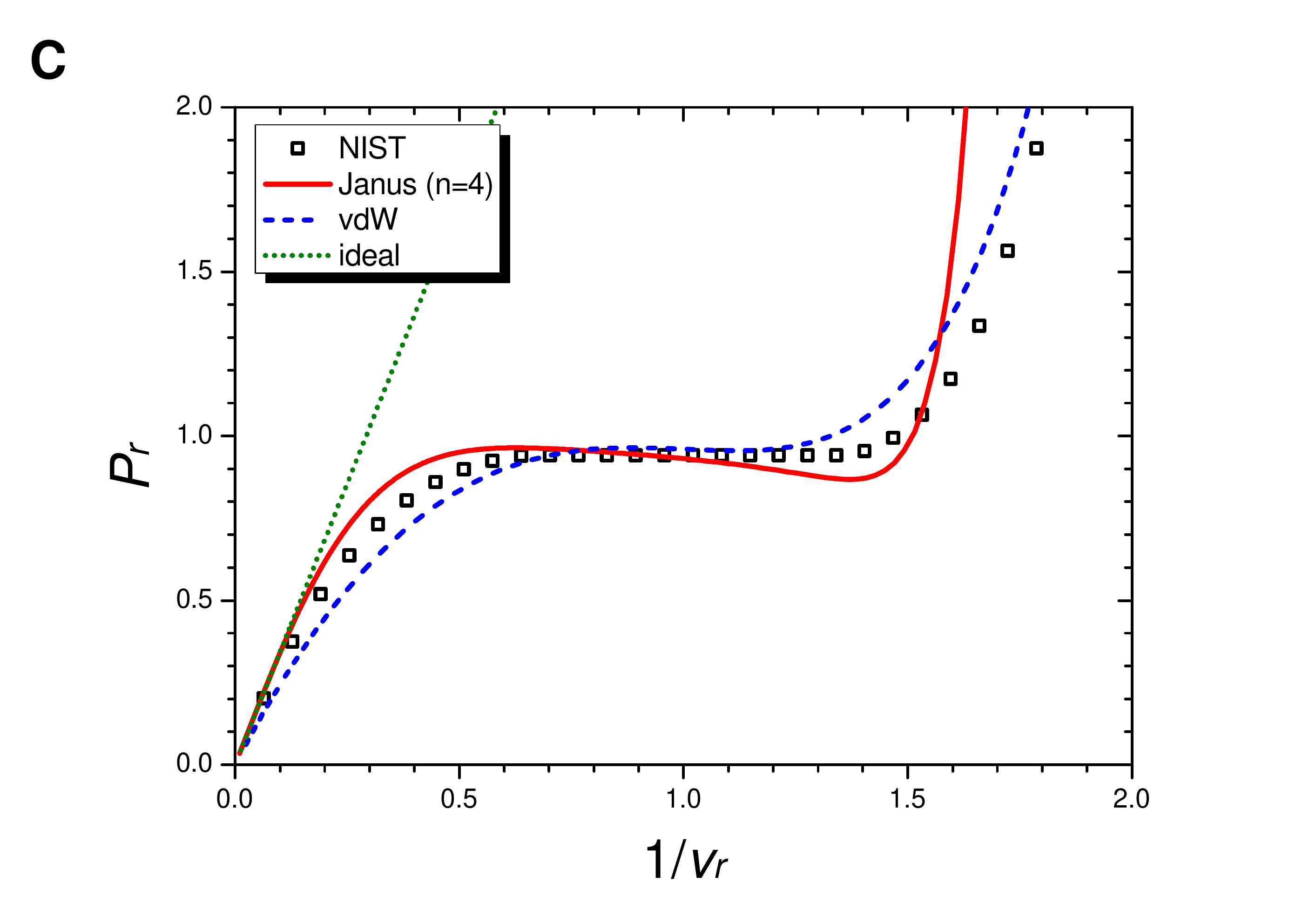}
		\caption{Isothermal  curves of nitrogen~($\mathrm{N}_{2}$)   at  $T_{r} = 1.01$  ({\bf{A}}), $T_{r} = 1.00$  ({\bf{B}}), and  $T_{r} = 0.99$ ({\bf{C}}). Boxes are from the NIST  data. The red solid line is drawn  from the $n=4$ Janus van der Waals equation~(\ref{n4JvdW}) and is better fitted than   the original van der Waals equation (\ref{vdW}) or the classical ideal gas law~(\ref{ideal}).}\label{nitrogen_isotherm}
	\end{center}
\end{figure}

\textbf{Figure~\ref{nitrogen_isotherm}}  shows the  isothermal  curves  of   nitrogen~($\mathrm{N}_{2}$) at $T_{r} = 1.01$  ({\bf{A}}), $T_{r} = 1.00$ ({\bf{B}}), and  $T_{r} = 0.99$  ({\bf{C}})  respectively. They are 
drawn by the NIST data and further by the three equations: the  $n=4$ Janus van der Waals equation~(\ref{n4JvdW}), the original van der Waals equation (\ref{vdW}), and the classical ideal gas law (\ref{ideal}).  The Janus van der Waals equation shows enhanced sigmoid shape compared to the original van der Waals equation when $T_{r}$ is lower than $1$. % When $T_{r} \ge 1$,  the Janus van der Waals equation fits very well with the NIST data near the  critical point. 

\newpage

\begin{figure}[H]
	\begin{center}
		\includegraphics[width=17.0cm]{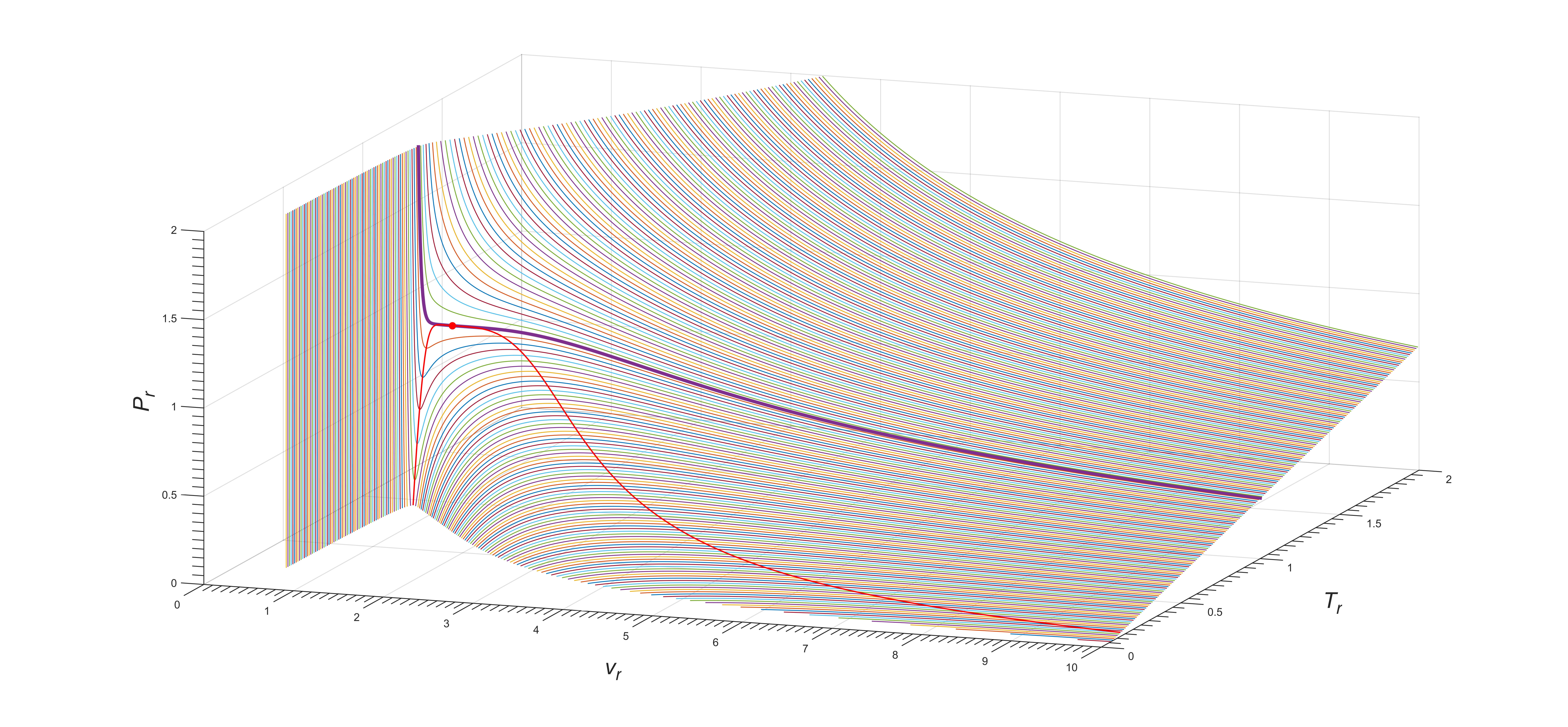}
		\caption{Three-dimensional $P_{r}-v_{r}-T_{r}$ phase diagram of the exact  $n=4$  Janus van der Waals equation~(\ref{bcvdW}), as for nitrogen  ($\mathrm{N}_{2}$).  The bold purple line corresponds to the isotherm of $T_{r} = 1.00$ as depicted in  \textbf{Figure~\ref{nitrogen_isotherm}~{\bf{B}}}; the red line is the Janus van der Waals spinodal curve with $a = 0.99$;  and the red dot is the critical point.}\label{nitrogen_PVT_spinodal}
	\end{center}
\end{figure}

\textbf{Figure~\ref{nitrogen_PVT_spinodal}}  shows  the three-dimensional $P_{r}-v_{r}-T_{r}$ phase diagram of the  exact  $n=4$   Janus van der Waals equation~(\ref{bcvdW}) with (\ref{fform}), as for  nitrogen  ($\mathrm{N}_{2}$).  The red line shows  the spinodal curve~(\ref{spinodalT}), \textit{i.e.~}$\partial_{v}P(T,v)=0$,  of the Janus van der Waals equation with the choice of $a = 0.99$. 

\section{Discussion}
Our Janus van der Waals equations  may appear somewhat similar to other known equations of state,  such as the virial equation of state, a typical example  of series expansion~\cite{virial}. Irrespective of the similarity of appearance to other equations, the  Janus van der Waals equations  differ totally from  others, as the starting point  is  directly from statistical physics  itself at quantum level: namely that the partition function of any finite system  should be analytic. Alternative to the conventional thermodynamic limit~\cite{YangLee,Kadanoff},   if we  persistently take  the analyticity of a partition function  for  granted~\cite{7616PRA,7616relativistic,Jeon:2011nk,Park:2013gpa,Cho:2014joa}, it becomes   the spinodal  curve itself that   draws the liquid-gas phase diagram where  a critical point is  identified as the extremum of the spinodal  curve, see \cite{7616relativistic} and Figure 6 therein.  One surprising result of the previous work by two of us~\cite{Cho:2016jzz} was the possibility of having  more than one critical points which should be very close to each other. Our ansatz  (\ref{spinodalT})  modifies   the original van der Waals equation and realises  the idea of  multi-critical points in a simple manner,   restricted to even critical indices, ${n_{c}=n=2,4,6}$ for ${T<T_{c}}$ and ${n_{c}=2}$ for ${T>T_{c}}$, or following the notation of \cite{Cho:2016jzz}, $(n_{+},n_{-})=(2,n)$.  Our proposed van der Waals equations  then naturally explain the two-sided phase transitions reported in \cite{Cho:2016jzz} and provide overall  effective  descriptions of real molecules, in particular better  than the original van der Waals equation as well as  the classical ideal gas law.

\begin{figure}[H]
	\begin{center}
		\includegraphics[width=17cm]{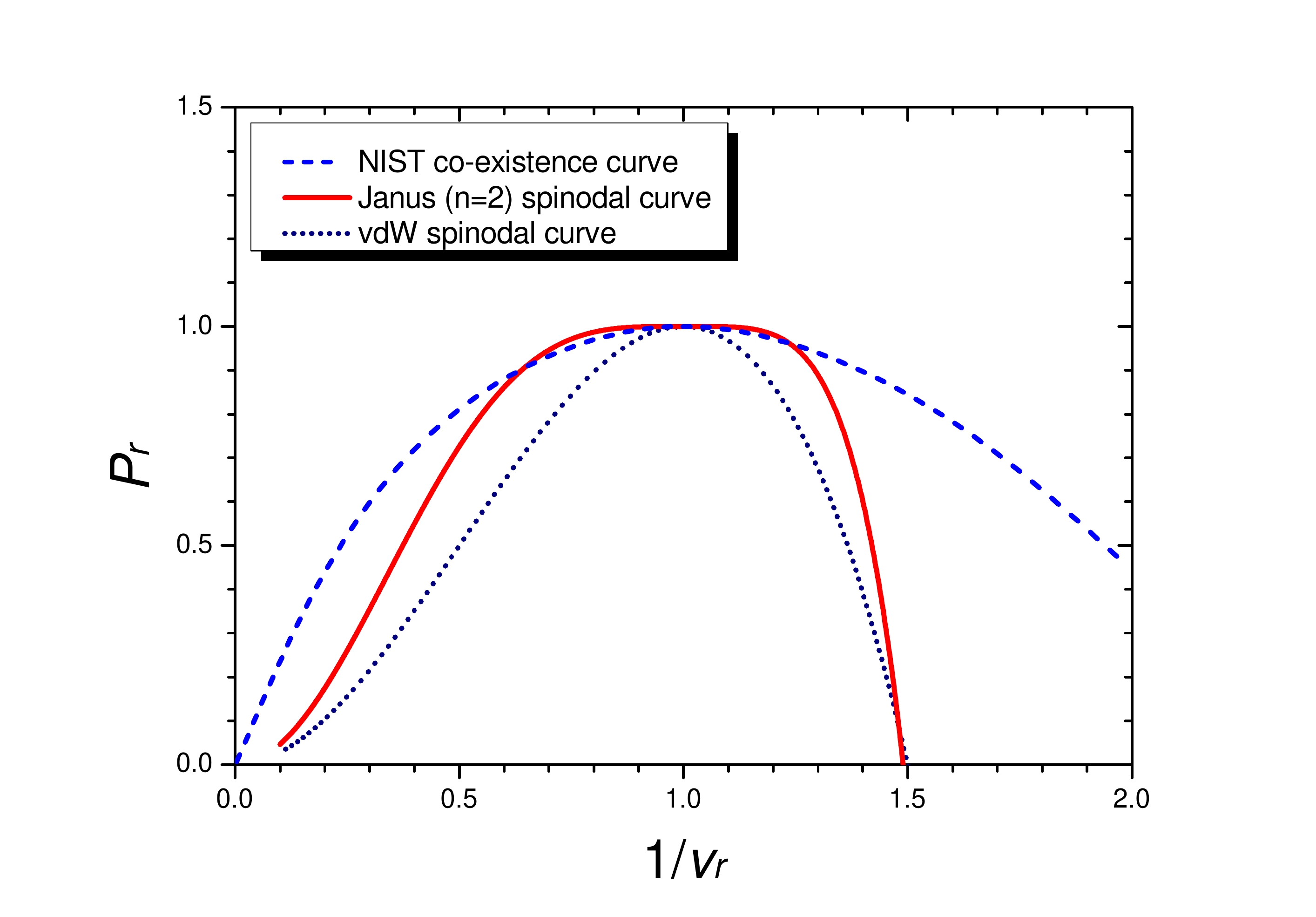}
		\caption{Comparison of the NIST co-existence curve data of cyclopentane ($\mathrm{C_{5} H_{10}}$)  \blue{(blue colored)} with the  Janus van der Waals spinodal curve  \red{(red colored,   $a = 0.99$)}  as well as  with  the original van der Waals spinodal curve (dotted black  line).  It is a conjecture of Ref.\cite{7616relativistic} that the liquid-gas co-existence curve should  actually  coincide with the spinodal curve. Improving our proposed Janus van der Waals equations further, this may be realisable.}\label{cyclopentane_spinodal_coexistence}
	\end{center}
\end{figure}

In \textbf{Figure~\ref{cyclopentane_spinodal_coexistence}}, we have compared the  NIST co-existence curve data of cyclopentane molecule ($\mathrm{C_{5} H_{10}}$)   with the   Janus van der Waals spinodal curve~(\ref{spinodalT}) of ${n=2}$,  ${a = 0.99}$ and also with  the original van der Waals spinodal curve~(\ref{spinodalvdW}).  The Janus van der Waals spinodal curve fits well   the NIST co-existence curve data in a wider range near the critical point,  though not perfect.  Further modifications of the present Janus van der Waals equations to  match the spinodal curves with the co-existence curves of real molecules will lead to more realistic, improved  equations of state.  Such modifications may  require  more than two critical points,   generalising the ansatz  (\ref{spinodalT}):
\be
T_{r}=-(v_{r}-b)^{2}\frac{\rd f_{\vec{n}}(v_{r})}{\rd v_{r}}=1\,-\,\prod_{i=1}^{N}\frac{(v_{r}-a_{i})^{n_{i}}}{v_{r}}\,.
\ee
Here  $n_{i}$'s are natural numbers,  and  especially those  $a_{i}$'s with $n_{i}\geq 2$ (even as well as odd) correspond to  multi-critical points. The largest value of such $a_{i}$'s should be exactly  unity with the critical index $2$ as the NIST data suggests~\cite{Cho:2016jzz}, while the smallest one   should be still  close to unity. Further, the former should be a local maximum of $T_{r}$, while  the latter should be either a local maximum as in  \textbf{Figure~\ref{FIG1}} if  the critical index is even  or   an inflection point  if it is odd.   We leave  the construction of this kind of multi-critical  Janus van der Waals equations  for future work.

The three Janus van der Waals equations~(\ref{n2JvdW}), (\ref{n4JvdW}), (\ref{n6JvdW}) have been obtained after taking the limit $a\rightarrow 1^{-}$. Thus, the formulas should not be used to see the two-sided critical phase transitions for which  the exact formula (\ref{bcvdW}) with (\ref{fform}) must be taken and   zoomed  in sufficiently.    When zoomed out, or moderately away from the two critical points,  the two  powers, $(v_{r}-a)^{n}$ and $(v_{r}-1)^{2}$ in (\ref{spinodalT}), may appear   converging to $(v_{r}-1)^{n+2}$ and mimic     an enhanced critical index $n_{c}=n+2$.  This implies the critical exponents   $\alpha_{\scriptscriptstyle{P}}=\gamma_{\scriptscriptstyle{P}}
=\frac{n+2}{n+3}$, $\beta_{\scriptscriptstyle{P}}=\delta^{-1}=\frac{1}{n+3}$ and also explains the `flatness' of the top of the spinodal curve in  \textbf{Figure~\ref{cyclopentane_spinodal_coexistence}}. The  three    formulas (\ref{n2JvdW}), (\ref{n4JvdW}), (\ref{n6JvdW}) are for such  effective descriptions.   The NIST data analyses of \cite{Cho:2016jzz}, in particular the figures~2,3,4 therein, seem to agree   with  this enhancement moderately  away from the critical points.

Having the Janus van der Waals equations  completely determined, it is worth while to   recall  
\be
P_{r}=\chi T_{r}\frac{\partial \ln Z(T_{r},v_{r})}{\partial v_{r}}\,,
\ee
and to  obtain the underlying  partition function (per particle),
%%
%\small{\be
%\ln Z=\ln(v_{r}-b)+\frac{1}{\,b^{4}T_{r}}\sum_{l=0}^{n+1}\,\frac{c_{l}}{(n+3-l)(n+2-l)}\left(\frac{b}{v_{r}}\right)^{n+2-l}
%\ee }
%% 
\be
\ln Z=\ln\!\left[(v_{r}{-b})T_{r}^{3/2}\right]+{\dis{\sum_{l=0}^{n+1}}}\,\frac{c_{l}({b}/v_{r})^{n+2-l},}{\,(n+3-l)(n+2-l)b^{4}T_{r}}\,,
\ee
%%%
%\small{\be
%\!\ln Z=\ln\!\left[(v_{r}-b)T_{r}^{3/2}\!\right]+\sum_{l=0}^{n+1}\,
%\frac{c_{l}}{\,2\binom{n+3-l}{2}b^{4}T_{r}}\left(\frac{b}{v_{r}}\right)^{n+2-l},
%\ee} 
%%%
where the  constant of integration $\frac{3}{2}\ln T_{r}$ has been added to ensure  the  isochoric specific heat   $c_{v}=\frac{3}{2}\kB$ at high temperature.

Given the good agreement of    the Janus van der Waals equations and  the NIST Reference Data, which we report in this work, we call for further investigation of the multi-critical points  and   the analyticity of  partition functions questioning the (rather dogmatic) thermodynamic limit.

\section*{Conflict of Interest Statement}
The authors declare that the research was conducted in the absence of any commercial or financial relationships that could be construed as a potential conflict of interest.

\section*{Author Contributions}
J-HP proposed the research and derived the formulas. JK contributed the NIST Reference Data handling and analysis. D-HK led the interpretation of the data.  D-HK and J-HP wrote the manuscript.

\section*{Funding}
This work was  supported by  Basic Science Research Program through the National Research Foundation of Korea Grants, NRF-2016R1D1A1B01015196
and NRF-2020R1A6A1A03047877 (Center for Quantum
Space Time).

\section*{Acknowledgment}
We thank  KyuHwan Lee for technical help at the early stage of the project.

\section*{Supplemental Data}
Figures of the eight molecules belonging to the $n=4$ case and of  helium-4 of the $n=6$ case are included in the Supplementary Material (attached after References).  

\section*{Data Availability Statement}
The datasets for this study  is 
\textit{NIST Reference Fluid Thermodynamic and Transport Properties Database (REFPROP): Version 9.1, National Institute of Standards and Technology (NIST), Standard Reference Data Program, Gaithersburg} and can be found at  \url{https://www.nist.gov/srd/refprop}~\cite{NIST}.

\newpage

\includepdf[pages=-]{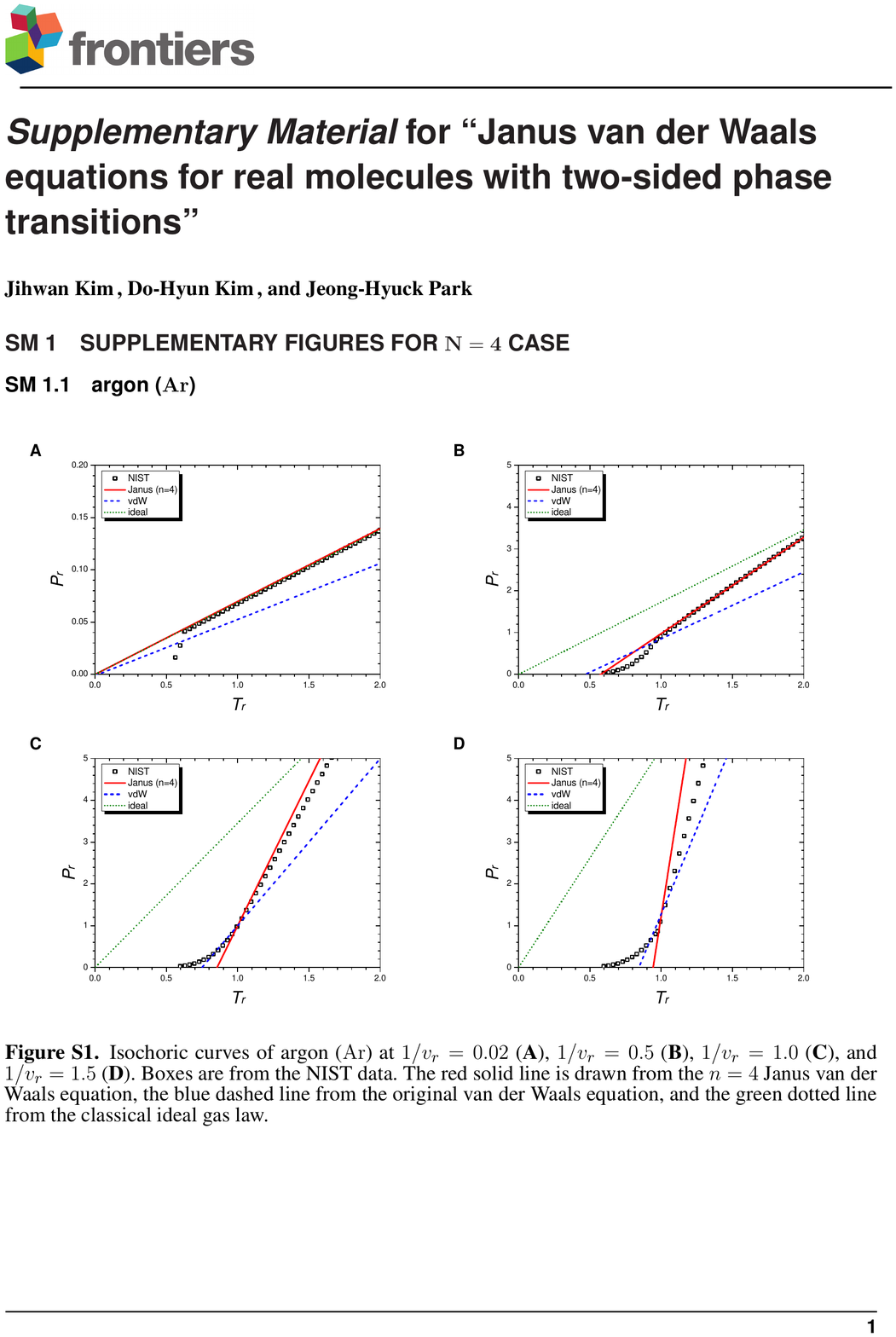}

\end{document}